\documentclass[12pt]{amsart}
\usepackage{amsfonts,amssymb,amsthm,amsmath}
\usepackage{latexsym}

\usepackage{epsfig}
\DeclareSymbolFont{boperators}{OT1}{cmr}{bx}{n}

\DeclareMathAccent{\karika}{\mathalpha}{boperators}{"17}

 \newcommand{\R}[1]{(\ref{#1})}
            \newcommand{\C}[1]{\cite{#1}}

            \newcommand{\mvec}[1]{\mbox{\boldmath{$#1$}}}
\newcommand{\e}{u}

\newcommand{\M}{{M}}
  \newcommand{\x}{\mvec{x}}

  \newcommand{\be}{\begin{equation}}
            \newcommand{\en}{\end{equation}}
\newcommand{\beg}{\begin{eqnarray}}
            \newcommand{\ene}{\end{eqnarray}}

  \newcommand{\m}{\mvec{m}}

\newcommand{\tr}{\mathop{\textnormal{tr}}\nolimits }

\renewcommand{\bold}[1]{\mathbf{#1}}

\renewcommand{\text}[1]{\mbox{#1}}

\begin{document}

\author { C.~Papenfuss, J.~Verh\'as, W.~Muschik}

\title {Thermodynamic Theory for Fiber Suspensions}

\address{C.~Papenfuss, W.~Muschik: Technische Universit\"at
Berlin \newline D-10623 Berlin, Germany }

\address{J.~Verh\'as:  Technical University of Budapest
\newline H-1521 Budapest, Hungary }

\date{\today }

\begin{abstract}

In this paper three different approaches towards a continuum
theory of fiber suspensions are discussed. The first one is the
classical Thermodynamics of Irreversible Processes with internal
variables. It derives constitutive equations for fiber suspensions
on the basis of ONSAGERs phenomenological coefficients, which  are
related to the mechanical properties of the fibers. Secondly
another method of exploiting the dissipation inequality, the
method introduced by LIU  is applied in order to derive results on
constitutive equations. Finally a mesoscopic background theory is
discussed. In this
 approach
a distribution of different fiber orientations and fiber
deformations is assumed. The fiber orientation and deformation are
additional variables in the domain of field quantities. The new
field quantities on the enlarged set of variables
 obey balance
 equations.  The mesoscopic balance of mass results in an equation of motion for the
distribution of fiber orientations and deformations. The usual
macroscopic fields of continuum mechanics are averages of
mesoscopic fields over the different orientations and
deformations. Macroscopic quantities characterizing the
orientational   order, the alignment tensors are defined by the
help of the orientation distribution function. They are internal
variables in the macroscopic theory characterizing the
distribution of fiber orientations.

The result of the Irreversible Thermodynamics approach mainly
concerns the stress tensor. Cauchy's stress is usually assumed to
be symmetric, but  in fiber suspensions it has an anti-symmetric
part, too. The skew-symmetric part of the stress shows up in  the
balance equation of angular momentum; it is related to the torque
density, the rate of internal moment of momentum, and the couple
stress. The latter is a result of the internal structure of the
material.
 The fibers are assumed to be
microscopic and the stress field smooth on macroscopic length
scale. If the local structure of the flowing medium may be
characterized by second order pseudo (axial) tensors as internal
variables  then in the linear Onsager equations there appears
couple stress coupled to the gradient of the angular velocity.

The results of the Liu-procedure are consistent with this
observation. The couple stress can be calculated as a combination
of derivatives of the free energy density (see equation
(\ref{pi_biax})), which may have an anti-symmetric part, too. This
result may be used to calculate the couple stress from an ansatz
for the free energy density,
 taking into account the deformation of fibers.

Another result of the Liu-exploitation of the dissipation
inequality is that the assumption of the entropy flux being heat
flux over temperature does not contradict the second law of
thermodynamics, but it does not necessarily follow from the
dissipation inequality.

\end{abstract}

\maketitle

\footnotetext{
\\[1ex]
This work has been supported by the Hungarian National
Scientific Research Funds, OTKA (T-17000) and
MKM (FKFP 0287/1997), the EC (Contract No: ICOP-DISS-2168-96), the DAAD, and the Draloric company (Selb, Germany).
}

\section{Introduction}
\setcounter{equation}{0}

A phenomenological theory of suspensions and emulsions of weakly
deformable spheres has been presented in \cite{Lhullier}. The
equations of motion for the particle deformation, and the
constitutive equation for the stress tensor have been derived in
the whole range of concentrations of the emulsion. They are
generalizations of  earlier work on dilute suspensions
\cite{Goddard, Roscoe} and emulsions \cite{Schowalter, Frankel}.
Thermodynamic arguments have been applied in order to reduce the
number of unknown constitutive coefficients. In semi-dilute
suspensions hydrodynamic interactions between particles are
important and their contribution to the stress tensor has been
investigated. As it is expected for all materials with an internal
structure the resulting stress tensor has an anti-symmetric part.
An overview over calculations of the stress tensor of suspensions
of elastic fibers of fixed orientation as well as on the stress
tensor of rigid orientable  fibers can be found in
\cite{Lhullier_CISM}.

A constitutive model taking into account the possible damage of fibers
of a glass-fiber woven polyester composite is presented in
\cite{Shen}. The material is viscoelastic with a dependency of elastic properties on the
micro-structural damage. The constitutive model is derived within the
framework of continuum mechanics and thermodynamics with an internal
variable. The phenomenological constitutive coefficients have been
obtained from experimental tests. Other experimental applications can
be found in \cite{Lou, Schapery, Schapery2}.

The derivation of constitutive equations on the ground of kinetic
theory is similar for polymers (see f. i. \cite{Doi_Edwards}) and
suspensions of fibers. An application of the stress tensor derived by
Batchelor \cite{Batchelor, Batchelor2} from a kinetic background theory to corner
flow of suspensions of orientable rigid rods can be found in
\cite{Hinch}.

Finally the methods applied in the present paper, namely continuum
mechanics with internal variables and Thermodynamics of Irreversible
Processes has been applied successfully in other fields: liquid
crystals \cite{Verhas2}, plasticity of metals and other solids
\cite{Rice}, polymer solutions. Composites with inextensible fibers
have been treated in \cite{rat} with continuum thermodynamics,
exploiting the dissipation inequality according to Liu \cite{Liu}.  In contrast to the present paper, where
the same method is applied in the second part, in \cite{rat} the material is a solid fiber composite.

In the last section the methods of the so called mesoscopic theory
are applied. This theory has been developed for liquid crystals
\cite{BLENK91,Journal,ZAMM92,PHYSICA}. For an overview over other
possible applications see \cite{Bam,poz}.

\section{Irreversible Thermodynamics of flexible fibers }
\setcounter{equation}{0}

The thermodynamic description of systems with dynamic (internal)
variables has been applied with success to several
kind of physical phenomena such as electric conduction
\cite{20}
electric and magnetic polarization
\cite{21, 22, 23}
heat conduction and radiation
\cite{3, 4, 5, 6, 7, 8}
viscoelastic and plastic deformations
\cite{1, 2, 11, 12, 13, 16, 18, 19, 24, 29, 31, 41, 42, 44, 46},
liquid crystals
\cite{45}. The thermodynamic theory of
visco-elasticity is based on dynamic variables which
are second order symmetric tensors. They are
related to the deformation of elastic micro-particles
or to the distribution function of the
orientation of rod like or disk like rigid
particles. For example some proteins, certain  saccharoides, and liquid crystals
 are well approximated by the above model, but it fails,
if particles are like elastic slender bars, e.g.,
in the case of  fiber suspensions, in
which the particles may be twisted and bent; their
affine
deformation is not relevant.

In the first part the flow induced anisotropy is neglected,
and the orientation distribution of the fibers is presumed
isotropic. The purpose here is to investigate the
consequences of the flexibility of the particles.

The deformation of a particle (bend and twist)
stores energy, so  influencing the
macroscopic mechanical properties of the fluid.

\vspace*{0.5cm}

 In the following we will denote by the symbol $: $
the contraction
 of second order tensors over all indices, in components: $\mvec{A}:\mvec{B} = A_{ik}B_{ki}$.
 The scalar product between vectors $\mvec{a}$ and $\mvec{b}$ is denoted by $\mvec{a}\cdot\mvec{b}$.


\subsection{Deformation of a Fiber}
In our model the fibers are assumed to be straight if not loaded.
 Then one can chose a coordinate $s$ along this fiber orientation and an orthogonal
 tensor $\mvec{U}(s)$
describing  the distortion of the  fiber. The macroscopic angular
distortion tensor  defined by
\begin{equation}
\boldsymbol\varphi := {\boldsymbol U}^T\cdot {d{\boldsymbol
U}\over d s } \label{eq:2.1}
\end{equation}
takes into  account  the local deformation of the flexible fibers,
and is the average over all fiber orientations in the continuum
element. $s$ is the local coordinate along the fiber,
 and $\mvec{x}$ is the position of the continuum element. $s$ is only introduced in order to describe the local
 fiber deformation.

  The
tensor \(\boldsymbol\varphi\) is obviously skew-symmetric as
\({\boldsymbol {U}}\) is orthogonal,
\begin{equation}
{\boldsymbol U}^T \cdot {\boldsymbol U} =\boldsymbol\delta,
\label{eq:2.2}
\end{equation}
where \(\boldsymbol\delta\) is the unit tensor and
\begin{equation}
\frac{d{\boldsymbol U}^T}{d s}\cdot {\boldsymbol U} + {\boldsymbol
U}^T \cdot \frac{d{\boldsymbol U}}{d s} = 0 \label{eq:2.3} ,
\end{equation} i.e.
\begin{equation}
\left({\boldsymbol U}^T \cdot {d{\boldsymbol U}\over d s}
\right)^T+ {\boldsymbol U}^T \cdot {d{\boldsymbol U}\over d s} = 0
\label{eq:2.4}.
\end{equation}

This way, we introduce the angular distortion vector (the vector
invariant of the angular distortion tensor) as
\begin{equation}
\vec\varphi\times\boldsymbol\delta =\boldsymbol\varphi
\label{eq:2.5}.
\end{equation}
Only in this case we will denote the vector by the symbol $\vec{\
}\ $ in order to distinguish
 it from the tensor $\mvec{\varphi}$.

Let \(\bold n\) denote the  unit vector tangential to the
undeformed fiber. The scalar product of \(\vec\varphi\) and
\(\bold n\) gives the twist:
\begin{equation}
t =\vec\varphi \cdot \bold n \label{eq:2.6},
\end{equation}
and the component of \(\vec\varphi\) perpendicular to \(\bold n\)
is the bend:
\begin{equation}
\bold b =\vec\varphi -\bold n (\bold n \cdot \vec\varphi)
\label{eq:2.7}
\end{equation}
As the fibers are elastic, two material coefficients $\mu_t$ and
$\mu_b$ have to be introduced to describe their stiffness. The
torque acting in a cross section  of the fiber is given by
\begin{equation}
\mvec{\tau} =\mu_t\bold n (\bold n\cdot\vec \varphi) + \mu_b
[\vec\varphi -\bold n (\bold n\cdot \vec\varphi)] \label{eq:2.8}
\end{equation}
The coefficient \(\mu_t\) is the torsion stiffness, and \(\mu_b\)
is the bend stiffness of the fiber. The elastic energy stored per
unit length of a fiber is given by
\begin{equation}
u_e = {\mu_t\over2} (\vec\varphi\cdot \bold n)^2+
{\mu_b\over2}(\vec\varphi^2-(\vec\varphi\cdot \bold n)^2).
\label{eq:2.9}
\end{equation}
This way, we have obtained the expression for the energy stored by
a deformed fiber.

The fibers are deformed by flow and they continuously relax. The
angular distortion of an individual fiber \(\vec\varphi\) depends
on the local orientation of the fiber,
\begin{equation}
\vec\varphi =\vec\varphi(\bold n). \label{eq:2.10}
\end{equation}
The function is odd, \(\vec\varphi(-\bold n)=-\vec\varphi(\bold
n)\), according to the definition of \(\vec\varphi\) eq.
(\ref{eq:2.1}), as turning the fiber (changing $\vec\varphi(\bold
n)$ to $\vec\varphi(-\bold n)$) gives a - sign in the line element
$ds$.

An obvious approximation for the function \(\vec\varphi(\bold n)\)
is the first term of its expansion by spherical harmonics;
\begin{equation}
\vec\varphi(\bold n)=\boldsymbol\alpha \cdot \bold n,
\label{eq:2.11}
\end{equation}
where \(\boldsymbol\alpha \) is a second order pseudo tensor as
\(\vec\varphi\) is an axial vector. VERHAS \cite{J.V.Borkow-paper}
elaborated the theory based on this linear approximation and the
theory gives only one relaxation time, while the relaxation times
for  twist and bend are expected to differ. As it will be shown in
this section, the above shortcoming is ceased if the approximation
(\ref{eq:2.11})
  is improved with a third order term (with a traceless tensor
  $\mvec{\beta}$):
\begin{equation}
\vec\varphi(\bold n)=\boldsymbol\alpha\cdot \bold n+\bold n(\bold
n \cdot \boldsymbol \beta \cdot \bold n), \label{eq:2.12}
\end{equation}
which can be decomposed as
\begin{equation}
\vec\varphi(\bold n)= \alpha^0 \bold n+\boldsymbol \alpha^a \cdot
\bold n + \boldsymbol \alpha^d \cdot \bold n - \bold n(\bold n
\cdot  \boldsymbol \alpha^d\cdot  \bold n) +\bold n(\bold n \cdot
\boldsymbol T\cdot  \bold n), \label{eq:2.13}
\end{equation}
where \(\alpha^0=1/3\tr\alpha\), \(\boldsymbol \alpha^a\) is the
skew symmetric part of \(\boldsymbol \alpha\), $\mvec{\alpha}^d$
is the deviatoric part of $\mvec{\alpha}$ and \(\boldsymbol T\) is
the deviatoric part of the sum of the tensors \(\boldsymbol
\alpha\) and \(\boldsymbol \beta\). The latter form shows the
twist and bend separately; the first and the last term are due to
twist, while the others are due to bend.

Inserting  the approximation equation (\ref{eq:2.12}), the elastic
energy of a fiber per unit fiber length reads
\begin{equation}
u_e (\mvec{n})={\mu_t \over 2}( \alpha^0 +\bold n \cdot
\boldsymbol T \cdot \bold n)^2 + {\mu_b \over 2}(\boldsymbol
\alpha^a \cdot \bold n + \boldsymbol \alpha^d \cdot \bold n -
\bold n(\bold n\cdot \boldsymbol \alpha^d \cdot \bold n))^2,
\label{eq:2.14}
\end{equation}
the average  of which  over all possible fiber orientations  is
\begin{eqnarray}
\lefteqn{u_{em}={1\over 4\pi}\oint_{S^2} u_e( \mvec{n}) d^2n =}
\nonumber\\&& ={\mu_t \over 2}\left({\alpha^0}^2+{2\over 15}
\boldsymbol T\boldsymbol : \boldsymbol T\right)+ {\mu_b \over
2}\left( {1\over3} \boldsymbol\alpha^a \boldsymbol :\boldsymbol
\alpha^a +{1\over5}\boldsymbol \alpha^d \boldsymbol : \boldsymbol
\alpha^d \right) \label{eq:2.15}
\end{eqnarray}
if uniform distribution of the fiber orientations is supposed,
i.e. the fiber orientations are distributed isotropically. A more
refined description introducing a non-uniform distribution of
orientations is discussed in the  third chapter.

If the total length of the fibers in a unit volume is denoted by
\(l\), we obtain
\begin{equation}
u_{nd}=lu_{em} \label{eq:2.16}
\end{equation}
for the part of the specific internal energy stored in the
deformation of the fibers, and not being dissipated. From the
formula, one can see that this not dissipated part of the energy
is a homogeneous quadratic function of the scalar \(\alpha^0\),
and the tensors \(\boldsymbol\alpha^a\), \( \boldsymbol\alpha^d\),
and \( \boldsymbol T\).

\subsection{Balance equations}
The balance equation for  mass is
\begin{equation}
{d\rho\over dt} +\rho\nabla \cdot \bold v = 0, \label{eq:3.1}
\end{equation}
which, for volume preserving motions (incompressible liquids),
reduces to
\begin{equation}
\nabla \cdot \bold v = 0. \label{eq:3.2}
\end{equation}
The derivative \( d\over dt\) is the material (or total) time
derivative
\begin{equation}
{d\over dt} = {
\partial\over
\partial t} + (\bold v\cdot \nabla) .
\label{eq:3.3}
\end{equation}
The balance equation of linear momentum is the usual one
\begin{equation}
\rho {d\bold v\over dt} =\rho\bold f +\nabla \cdot  \boldsymbol t
\label{eq:3.4}
\end{equation}
where \(\boldsymbol t\) is CAUCHYs stress tensor, \(\bold f\) is
the body force per unit mass, \(\rho \) is the density, and
\(\bold v\) is the velocity of the material. The definition for
the divergence used here reads in components: \be  (\nabla \cdot
\boldsymbol t)_i = \nabla_j t_{ij} \quad .\en

Although the internal angular momentum and the torque will be
neglected later, we write down the balance for the angular
momentum in general \cite{40, 47}.
\begin{equation}
\rho {d\,{\bold s}\over dt}= 2\bold w(\boldsymbol t) +\rho\bold m
+\nabla \cdot \boldsymbol\Pi \label{eq:3.5}
\end{equation}
Here \({\bold s}\) is the internal angular momentum per unit mass,
\(\bold m\) stands for the torque exerted on the material by
external fields and \(\boldsymbol\Pi \) is the couple stress
tensor. The vector \(\bold w(\boldsymbol t)\) stands for the
vector invariant of Cauchy's stress tensor. In the following
 we suppose that the average rotation of the fibers
(in addition to  thermal fluctuations)
is generated by the macroscopic mechanical motion.
Consequently, the internal angular  momentum (due to thermal motion) can be neglected compared to the moment
of momentum (due to macroscopic motion). Moreover, the rotational energy of internal angular momentum
need not be regarded. If we  also neglect the torque, which is reasonable as long as no
electro-magnetic fields are present, equation
(\ref{eq:3.5}) reduces to
\begin{equation}
2\bold w(\boldsymbol t) +\nabla \cdot \boldsymbol\Pi = 0
\label{eq:3.6}
\end{equation}
The balance equation of  internal energy is of basic importance
for constitutive modelling. For the total energy, we can write
\begin{eqnarray}
{d\over dt}\int \rho (u+{\bold v^2\over2}\,)dV= -\oint\bold q
\cdot d\bold A+ \oint\bold v \cdot\boldsymbol t\cdot d\bold A+
\nonumber\\
+\oint\mvec{\omega}_{\Pi}\cdot \boldsymbol\Pi\cdot d\bold A+
\int\rho\bold v \cdot \bold f\,dV+
\int\rho\mvec{\omega}_m\cdot \bold m\,dV
\label{eq:3.7}
\end{eqnarray}
The left hand side is the change of the total (internal plus kinetic) energy of the part of the
material regarded. Kinetic energy due to particle rotations has already been neglected. The terms on the
right hand side are heat flow, and power of stress, of couple stress, of
forces, and of couple forces.
The vectors \(\mvec{\omega}_\Pi \) and \(\mvec{\omega_m}\) are the angular
velocities of what the couple stress and the torque act on, respectively.

Now, we assume that both \(\mvec{\omega}_m\) and
\(\mvec{\omega}_\Pi \) are equal to the macroscopic angular
velocity of the body,
\begin{equation}
\mvec{\omega}_m=\mvec{\omega}_\Pi =\mvec{\omega }= {1\over2}\nabla
\times  \bold v. \label{eq:3.8}
\end{equation}
Making use of the other balance equations and
of  Gauss' theorem, we get
\begin{equation}
\rho {du\over dt} +\nabla \cdot  \bold q =\karika {\boldsymbol
d}\boldsymbol :\boldsymbol t^d+\boldsymbol\Pi\boldsymbol :\nabla
\mvec{\omega} \label{eq:3.9}
\end{equation}
where \(\karika{\boldsymbol d}\)  and  \(\boldsymbol  t^d\)
stand for the symmetric part of the velocity gradient and of
Cauchy's stress tensor, respectively.

\subsection{Entropy}

The independent state variables in the domain of the entropy  are
the internal energy, the pseudo scalar \(\alpha^0\) and the
tensors \(\boldsymbol \alpha^a\), \(\boldsymbol \alpha^d\) and
\(\boldsymbol T\); the latter taking into account  the state of
the deformation of the fibers.

Assume that the entropy depends on the dissipated energy only,
\begin{equation}
\eta  = \eta^e\big(u - {1\over\rho}u_{nd}\big). \label{eq:4.1}
\quad ,\end{equation} where $u_{nd}$ is the non-dissipated energy
(stored in the deformation of the fibers). From here and from
equations (\ref{eq:2.15}) and (\ref{eq:2.16}) we get
\begin{equation}
\eta=\eta^e\left(u - {l\mu_t\over 2\rho} \big[{\alpha^0}^2-{2\over
15} \boldsymbol T \boldsymbol : \boldsymbol T\big]-
{l\mu_b\over2\rho} \big[{1\over3} \boldsymbol\alpha^a \boldsymbol
: \boldsymbol \alpha^a +{1\over5}\boldsymbol \alpha^d \boldsymbol:
\boldsymbol \alpha^d \big]\right). \label{eq:4.2}
\end{equation}

The balance equation for the entropy, for vanishing entropy
supply,  is
\begin{equation}
\rho {d\eta\over dt} + \nabla \cdot \mvec{\phi} =\sigma \ge 0
\label{eq:4.3}
\end{equation}

The entropy flux is supposed to obey the constitutive equation for
local equilibrium systems
\begin{equation}
\mvec{\phi}={1\over T}\mvec{q} \qquad\text{where}\qquad {1\over T}
={
\partial s\over
\partial u} .
\label{eq:4.4}
\end{equation}
Making use of the entropy function in equation  (\ref{eq:4.2}) and
the balance equation of the internal energy, we obtain the
expression of the entropy production density
\begin{eqnarray}
\sigma &=&{1\over T}\left\{\karika{\boldsymbol d} \boldsymbol
:\boldsymbol t^d +\boldsymbol\Pi\boldsymbol :\nabla\mvec{\omega
}-l\mu_t\alpha^0\dot{\alpha^0} -{2l\mu_t\over15}\boldsymbol
T\boldsymbol :\karika{\boldsymbol T}- \vphantom{1\over T}\right.
\nonumber\\ &&\qquad
\left.-{l\mu_b\over3}\boldsymbol\alpha^a\boldsymbol:
\karika{\boldsymbol\alpha^a}
-{l\mu_b\over5}\boldsymbol\alpha^d\boldsymbol :
\karika{\boldsymbol\alpha^d} - {1\over T} \bold q \cdot\nabla
T\right\}. \label{eq:4.5}
\end{eqnarray}
Here \(\karika{\boldsymbol T}\), \(\karika{\boldsymbol\alpha^a}\),
and \(\karika{\boldsymbol\alpha^d}\) are the co-rotational time
derivatives of the tensors \(\boldsymbol T\),
\(\boldsymbol\alpha^a\), and \(\boldsymbol\alpha^d\),
respectively. The introduction of co-rotational derivatives is
here arbitrary, because it holds
\begin{eqnarray}
\mvec{T} :\karika{\mvec{T}} = \mvec{T} :\dot{\mvec{T}}\\
\mvec{\alpha^a} :\karika{\mvec{\alpha^a}} = \mvec{\alpha^a} :\dot{\mvec{\alpha^a}}\\
\mvec{\alpha^d} :\karika{\mvec{\alpha^d}} = \mvec{\alpha^d}
:\dot{\mvec{\alpha^d}} \quad .\end{eqnarray} It is motivated by
the fact that later fluxes and forces should be expressed in terms
of co-rotational time derivatives, i.e. the time derivatives in a
co-moving frame:
\begin{displaymath}
\karika{\boldsymbol T}= {d\boldsymbol T\over dt} +\boldsymbol
T\cdot\boldsymbol\omega - \boldsymbol\omega\cdot\boldsymbol T ,
\quad \karika{\boldsymbol \alpha^a}= {d\boldsymbol \alpha^a\over
dt} +\boldsymbol \alpha^a\cdot\boldsymbol\omega -
\boldsymbol\omega\cdot\boldsymbol \alpha^a , \quad
\karika{\boldsymbol \alpha^d}= {d\boldsymbol \alpha^d\over dt}
+\boldsymbol \alpha^d\cdot\boldsymbol\omega -
\boldsymbol\omega\cdot\boldsymbol \alpha^d ,
\end{displaymath}

The tensors $\boldsymbol\Pi$ and $\boldsymbol \nabla
\mvec{\omega}$ are decomposed into their symmetric and
anti-symmetric parts:
\begin{equation}
\boldsymbol\Pi\boldsymbol :\nabla\mvec{\omega}=
\boldsymbol\Pi^d\boldsymbol :
(\nabla\mvec{\omega})^d+\boldsymbol\Pi^a\boldsymbol :
(\nabla\mvec{\omega})^a. \label{eq:4.6}
\end{equation}

It results the entropy production density
\begin{eqnarray}
\lefteqn{T\sigma_s=\karika{\boldsymbol d} \boldsymbol :
\boldsymbol t^d +\boldsymbol\Pi^d\boldsymbol :
(\nabla\mvec{\omega})^d +\boldsymbol\Pi^a\boldsymbol :
(\nabla\mvec{\omega})^a -l\mu_t\alpha^0\dot{\alpha^0}} \nonumber\\
&&\qquad -{2l\mu_t\over15}\boldsymbol T\boldsymbol
:\karika{\boldsymbol T}
-{l\mu_b\over3}\boldsymbol\alpha^a\boldsymbol:
\karika{\boldsymbol\alpha^a}
-{l\mu_b\over5}\boldsymbol\alpha^d\boldsymbol :
\karika{\boldsymbol\alpha^d} -\bold q \cdot{1\over T}\nabla T.
\label{eq:4.7}
\end{eqnarray}
The first term on the right hand side refers to affine
deformations, the second and the third to distortions, the next
terms to the relaxations of the distortions of the fibers while
the last term is due to heat propagation. The set of process rates
is
\begin{displaymath}
\Big\{ \bold q;\ \boldsymbol t^d;\ \boldsymbol\Pi^d;\
\karika{\boldsymbol\alpha^d} ;\ \karika{\boldsymbol T};\
\boldsymbol\Pi^a;\ \karika{\boldsymbol\alpha^a};\ \dot{\alpha^0}
\Big\}
\end{displaymath}
and the set of corresponding forces is
\begin{eqnarray*}
\lefteqn{ \Big\{ -{1\over T}\nabla T; \karika{\boldsymbol d};\
(\nabla\mvec{\omega})^d;\ -{l\mu_b\over5}\boldsymbol\alpha^d;\
{2l\mu_t\over15}\boldsymbol T;}
\\&&\qquad
(\nabla\mvec{\omega})^a;\ -{l\mu_b\over3}\boldsymbol\alpha^a;\
-l\mu_t\alpha^0;\ \Big\}
\end{eqnarray*}

\subsection{Constitutive Equations}
In an isotropic material, with the above thermodynamic fluxes and
forces ONSAGERs linear laws read (\cite{9, 10, 15, 25, 32, 33, 34,
35, 39} )
\begin{eqnarray}
&\bold q = -{1\over T}L^q\nabla T=-\lambda\nabla T,&
\label{eq:5.1}
\\
&\boldsymbol t^d= 2\eta\karika{\boldsymbol d},& \label{eq:5.2}
\\[1em]
\boldsymbol\Pi^d&= L^d_{00}(\nabla\mvec{\omega})^d
-L^d_{01}{l\mu_b\over5}\boldsymbol\alpha^d -L^d_{02}
{2l\mu_t\over15}\boldsymbol T ,&
\nonumber\\
\karika{\boldsymbol\alpha^d}&= L^d_{10}(\nabla\mvec{\omega})^d
-L^d_{11}{l\mu_b\over5}\boldsymbol\alpha^d
-L^d_{12}{2l\mu_t\over15}\boldsymbol T \label{eq:5.3}
\\
\karika{\boldsymbol T}&= L^d_{20}(\nabla\mvec{\omega})^d
-L^d_{21}{l\mu_b\over5}\boldsymbol\alpha^d
-L^d_{22}{2l\mu_t\over15}\boldsymbol T
\nonumber\\[1em]
&\boldsymbol\Pi^a = L_{00}^a(\nabla\mvec{\omega})^a
-L^a_{01}{l\mu_b\over3}\boldsymbol\alpha^a,&
\nonumber\\
&\karika{\boldsymbol\alpha^a}= L^a_{10}(\nabla\mvec{\omega})^a
-L^a_{11}{l\mu_b\over3}\boldsymbol\alpha^a,& \label{eq:5.4}
\\[1em]
&{d\over dt}\alpha^0=-L^0l\mu_t\alpha^0.& \label{eq:5.5}
\end{eqnarray}

The pseudo scalar \(\alpha^0\) and the tensors
\(\boldsymbol\alpha^d\), and \(\boldsymbol T\) are related to the
distortion of the fibers, so we suppose they are of
\(\alpha\)-type, i.e., not changing under time inversion. On the
other hand \(\mvec{\omega} \) is of \(\beta\)-type, i.e., changing
sign under time inversion. The ONSAGER-CASIMIR  reciprocal
relations are
\begin{equation}
L^d_{10}= - L^d_{01}, \qquad L^d_{20}= - L^d_{02}, \qquad
L^d_{12}=L^d_{21}, \qquad L^a_{10}= - L^a_{01}. \label{eq:5.6}
\end{equation}
The inequalities
\begin{eqnarray}
\lambda >0,&\quad \eta >0,\quad L^d_{00}\ge 0,&\quad L^d_{11}\ge
0,
\nonumber\\
L^d_{22}\ge 0,&\quad L^a_{00}\ge 0,\quad L^a_{11}\ge 0,&\quad
L^0_{11}\ge 0
\label{eq:5.7}\\
&L^d_{11}L^d_{22}\ge (L^d_{12})^2& \nonumber
\end{eqnarray}
are the consequences of the Second Law of Thermodynamics.

According to CASIMIRs reciprocal relations, the Second Law gives
no restrictions  on the values of \(L^d_{01}\) and \(L^d_{20}\).
The general theory  does not imply any further restriction on the
value of the coefficients but the analysis of some particular
situations makes other simplifications obvious.

Equation (\ref{eq:5.5}) shows that the pseudo scalar  \(\alpha^0
\) tends to zero even if it were nonzero some time ago and it
cannot be excited at later times, because there is no coupling to
other variables. Therefore, after some relaxation time, \(\alpha^0
\)is always zero.

Because the trace of \(\boldsymbol\Pi\) does not appear in the
balance of
 energy and in the entropy production it is not relevant, and
 we conclude that  the
distinction between the symmetric and the deviatoric parts of the
tensor \(\boldsymbol\Pi\) is not necessary, i.e.,
\begin{displaymath}
\boldsymbol\Pi^s=\boldsymbol\Pi^d.
\end{displaymath}

There is no coupling between the symmetric part of the velocity
gradient (the affine deformation of the fluid) and the distortion
of the fibers because \(\karika d\) is the only second order polar
tensor among the thermodynamic forces. If we had regarded the
affine deformation of particles in the liquid, equation
(\ref{eq:5.2}) should have turned to the relation known in
visco-elasticity \cite{46},\cite{47}.

\subsection{Further considerations}

Up to here, the balance equations in general and the dissipation
inequality have been used only. Now we consider a special case, a
flow field which is such that the distortion of a previously
un-distorted fiber is the same as that of a fluid element, i.e.,
the equation
\begin{equation}
\karika{\vec\varphi}=\nabla\mvec{\omega}\cdot \bold n
\label{eq:6.1}
\end{equation}
holds if \(\boldsymbol \alpha^a = \boldsymbol \alpha^d =
\boldsymbol T=0\). Comparing it with equation (\ref{eq:2.13}) and
using equations (\ref{eq:5.3}) and (\ref{eq:5.4}) and
(\ref{eq:5.5}) we get that
\begin{eqnarray}
\lefteqn{ \nabla\mvec{\omega}\cdot \bold
n=L^a_{10}(\nabla\mvec{\omega})^a\cdot\bold n+
L^d_{10}(\nabla\mvec{\omega})^d \cdot \bold n-} \nonumber\\&&
-L^d_{10}\bold n \big[\bold n\cdot (\nabla\mvec{\omega})^d\cdot
\bold n\big] +L^d_{20}\bold n\big[\bold
n\cdot(\nabla\mvec{\omega})^d\cdot\bold n\big] \label{eq:6.2}
\end{eqnarray}
must  hold for any \(\bold n\). This equation is compared with the
general decomposition of the tensor $\nabla \mvec{\omega}$ into
its symmetric  part and antisymmetric part.  As the
phenomenological coefficients do not depend on the flow geometry,
we conclude that
\begin{equation}
L^a_{10}=L^d_{10}=1\qquad \text{and}\qquad L^d_{10}=L^d_{20}.
\label{eq:6.3}
\end{equation}
If we make the reasonable assumption  that the bend relaxes with a
single relaxation time and that the twist and the bend relax
separately, we get that
\begin{equation}
3L^d_{11}=5L^a_{11}\qquad\text{and}\qquad L^d_{12}=0.
\label{eq:6.4}
\end{equation}
These are relations between the constant (within the linear
theory) constitutive coefficients, and therefore they are valid
not only in the special case.

With these  results the  constitutive equations (\ref{eq:5.3}) get
the form
\begin{eqnarray}
&\boldsymbol\Pi=
L^d_{00}(\nabla\mvec{\omega})^d+L_{00}^a(\nabla\mvec{\omega})^a
-{l\mu_b\over5}\boldsymbol\alpha^d-{l\mu_b\over3}\boldsymbol\alpha^a
-{2l\mu_t\over15}\boldsymbol{T} ,&
\nonumber\\
&\karika{\boldsymbol\alpha}= (\nabla\mvec{\omega})
-L^d_{11}{l\mu_b\over5}\boldsymbol\alpha \label{eq:6.5}
\\
&\karika{\boldsymbol T}= (\nabla\mvec{\omega})^a
-L^d_{22}{2l\mu_t\over15}\boldsymbol T. \nonumber
\end{eqnarray}
The number of  material coefficients to be determined
experimentally has been reduced to six.

We have shown  that the stress tensor may have a skew symmetric
part in a fiber suspension. The equations of motion are those of a
micropolar continuum.
 Moreover, the theory based on the approximation in
equation (\ref{eq:2.13}) takes into account  the fact that the
twist and the bend of the fibers relax with different relaxation
times. The authors expect, that the improvement of the
approximation (\ref{eq:2.13}) to higher orders does not change the
equations because of the nature of the isotropic tensors. To
decide if it is really so needs further investigations.

\section{Exploitation of the dissipation inequality according to
LIU}\label{Liu} \setcounter{equation}{0} In this section we want
to derive the restrictions on constitutive functions for fiber
suspensions by the method of LIU \cite{Liu}.  From a very basic
amendment to the second law of thermodynamics on physical
arguments \cite{amendment} one can show that the requirement of a
positive entropy production density restricts possible
constitutive functions and does  {\em not} rule out certain
process directions in non-equilibrium. The most general way to
derive these restrictions on constitutive functions is the method
by LIU \cite{Liu}. The restrictions follow after the set of
variables for material properties, the state space, has been
chosen. They do not completely determine the material behavior, as
it is natural, because there are different materials with the same
set of variables, but different constitutive properties. The
results of this method give the most general framework compatible
with the second law of thermodynamics. In the example of fiber
suspensions we chose the state space

\be Z = \{ \rho , T, \mvec{v}, \nabla \mvec{v} , \mvec{Q},
\mvec{O}, \mvec{\Omega} \} \label{Z_biax} \quad , \en where all
constitutive quantities are assumed not to depend on $\mvec{v}$.
This state space is related  to the set of variables in the domain
of the constitutive functions entropy and internal energy in the
previous section. Apart from the equilibrium variables mass
density and temperature it includes the velocity gradient, an
orthogonal second order tensor $\mvec{Q}$ describing the local
fiber orientation (which is assumed variable here). $\mvec{Q}$ is
the mapping between the orientation vector of the (undistorted)
fiber and a reference coordinate system. With the tensor $Q$ it
would also be possible to account for the orientation of biaxial
elements, such as plates, but this is out of the scope of the
present paper. It is possible here, that there is an arbitrary
distribution of fiber orientations. Then the orientation
distribution function can be approximated by a second order
tensor,  denoted as $\mvec{A} $. The definition of this second
order tensor from the distribution function will be discussed in
section \ref{meso}. Then the tensor $\mvec{Q}$ is the mapping
between the coordinate axes  and the principal axes of the tensor
$\mvec{A}$.

We also include  the second order tensor $\mvec{O}$ for the local
deformation of the fibers. $\mvec{O}$ is defined as \be \mvec{O} =
\frac{1}{2} \left( (\nabla \mvec{Q}^T ) \cdot \mvec{Q} \right)
:\mvec{\epsilon} \en with the totally antisymmetric third order
tensor $\mvec{\epsilon}$.

In contrast to the previous section we also include the time
derivative $\Omega$ of the fiber orientation: \be \frac{\partial
\mvec{Q}}{\partial t} = \mvec{\Omega} \times \mvec{Q}  \en
 in the set
of relevant variables for constitutive functions.

As the materials discussed here are micropolar media we will take
into account the balance of internal angular momentum in addition
to the balances of mass, momentum, and energy. The following
inequality has to be exploited:

 \beg
\rho \frac{d \eta}{dt} + \nabla \cdot \mvec{\phi} - \frac{r}{T} +
\lambda^{\rho} \left( \frac{d \rho}{dt} + \rho \nabla \cdot
\mvec{v}\right)\nonumber \\ + \lambda^{p} \left( \rho \frac{d
\mvec{v}}{dt}- \nabla \cdot \mvec{t}- \rho\mvec{f}\right)
\nonumber \\+ \lambda^u \left( \rho \frac{d u}{dt} + \nabla \cdot
\mvec{q}- \mvec{t}:\nabla
 \mvec{v}- r +\mvec{\Pi} : \nabla  \mvec{\Omega}\right)\nonumber \\+ \mvec{\lambda}^s
 \cdot \left( \rho \frac{d\mvec{s}}{dt} - \mvec{\epsilon} : \mvec{t} + \nabla \cdot
 \mvec{\Pi}^T + \rho \mvec{m} \right)\geq 0 \label{Liu_biax}\quad .\ene
 After exploiting the differentiations of the constitutive
 functions, defined on the state space (\ref{Z_biax}), according to the chain rule it results an inequality
 linear in the following higher derivatives:
 \beg
\dot T, \dot \rho, \dot{ \mvec{v}}, \frac{d (\nabla
\mvec{v})}{dt}, \dot{\mvec{O}},   \dot{\mvec{\Omega}},  \nabla
\rho , \nabla T, \nabla \nabla \mvec{v}, \nabla \mvec{O}, \nabla
\Omega
  \quad . \ene

These higher derivatives are not all independent, but one
constraint between them has to be taken into account. To show this
we will use components with respect to a cartesian coordinate
system:

{\bf Proposition}:

\be \frac{d \mvec{O}}{dt}= \left( \nabla \mvec{\Omega} \right)^T +
\frac{1}{2} \left( \mvec{\epsilon} \cdot \mvec{Q} : \frac{\partial
\mvec{Q}^T}{\partial \mvec{x}}\right)\cdot \frac{\partial
\mvec{v}}{\partial \mvec{x}} \en or in components \be \frac{d
O_{ik}}{dt}=  \frac{\partial \Omega_i}{\partial x_k}  +
\frac{1}{2} \epsilon_{irl} Q_{ls}  \frac{\partial Q_{rs}}{\partial
x_m} \frac{\partial v_m}{\partial x_k} \label{constraint} \quad .
\en The proof of this proposition is shown in the appendix.
  $\mvec{Q}$,  $\nabla \mvec{v}$ and $\nabla
\mvec{Q} $ are state space variables, whereas $\nabla
\mvec{\Omega}$ is not included in the state space. This gradient
shows up in the list of higher derivatives.

After inserting this constraint we can write down the
LIU-equations, corresponding to the different higher derivatives:
\beg \dot{\rho}: \quad \rho \frac{\partial \eta}{\partial \rho} +
\lambda^u \rho \frac{\partial \e}{\partial \rho} +\lambda^{\rho} +
\rho \mvec{\lambda}^s \cdot \frac{\partial \mvec{s}}{\partial
\rho} =0
\label{dot_rho}\\
\dot{T}: \quad \rho \frac{\partial \eta}{\partial T} + \lambda^u
\rho \frac{\partial \e}{\partial T} + \rho \mvec{\lambda}^s \cdot
\frac{\partial \mvec{s}}{\partial T} =0
\label{dot_T}\\
\dot{\nabla \mvec{v}}: \quad \rho \frac{\partial \eta}{\partial
\nabla \mvec{v}} + \lambda^u \rho \frac{\partial \e}{\partial
\nabla \mvec{v}}  + \rho \mvec{\lambda}^s \cdot \frac{\partial
\mvec{s}}{\partial \nabla \mvec{v}} =0
\label{dot_nabla_v}\\
\nabla \mvec{\Omega} : \quad \rho \frac{\partial \eta}{\partial
\mvec{O}} + \frac{\partial \mvec{\Phi}}{\partial \mvec{\Omega}} -
\mvec{\lambda}^p \cdot \frac{\partial \mvec{t}}{\partial
\mvec{\Omega}} + \nonumber \\  \mvec{\lambda}^s\cdot \left(\rho
\frac{\partial \mvec{s}}{\partial \mvec{O}} - \frac{\partial
\mvec{\Pi}^T}{\partial \mvec{\Omega}} \right) + \lambda^u \left(
-\mvec{\Pi}+ \frac{\partial \mvec{q}}{\partial \mvec{\Omega}}\right) =0 \label{nabla_omega} \\
\mvec{\dot{\Omega}}: \quad \rho \frac{\partial \eta}{\partial
\mvec{\Omega}} + \lambda^u\frac{\partial \e}{\partial
\mvec{\Omega}} +  \rho \mvec{\lambda}^s\cdot \frac{\partial
\mvec{s}}{\partial \mvec{\Omega}}  = 0 \label{dot_omega}\\
\nabla \rho : \quad \frac{\partial \mvec{\Phi}}{\partial \rho} -
\mvec{\lambda}^p \cdot \frac{\partial \mvec{t}}{\partial \rho} -
\mvec{\lambda}^s \cdot \frac{\partial \mvec{\Pi}^T}{\partial \rho}
+ \lambda^u \frac{\partial \mvec{q}}{\partial \rho} = 0
\label{nabla_rho}\\
\nabla T : \quad \frac{\partial \mvec{\Phi}}{\partial T} -
\mvec{\lambda}^p \cdot \frac{\partial \mvec{t}}{\partial T} -
\mvec{\lambda}^s \cdot \frac{\partial \mvec{\Pi}^T}{\partial T} +
\lambda^u \frac{\partial \mvec{q}}{\partial T} = 0
\label{nabla_T}\\ \nabla \nabla \mvec{v} : \quad \frac{\partial
\mvec{\Phi}}{\partial \nabla \mvec{v}} - \mvec{\lambda}^p \cdot
\frac{\partial \mvec{t}}{\partial \nabla \mvec{v}} -
\mvec{\lambda}^s \cdot \frac{\partial \mvec{\Pi}^T}{\partial
\nabla \mvec{v}} + \lambda^u \frac{\partial \mvec{q}}{\partial
\nabla \mvec{v}} = 0 \label{nabla_nabla_v}\\ \nabla \mvec{O} :
\quad \frac{\partial \mvec{\Phi}}{\partial \mvec{O}} -
\mvec{\lambda}^p \cdot \frac{\partial \mvec{t}}{\partial \mvec{O}}
- \mvec{\lambda}^s \cdot \frac{\partial \mvec{\Pi}^T}{\partial
\mvec{O}} + \lambda^u
\frac{\partial \mvec{q}}{\partial \mvec{O}} = 0 \label{nabla_O}\\
 \frac{d \mvec{v}}{dt} : \quad \rho \mvec{\lambda}^p = 0
\label{dot_v} \quad . \ene
 From this set of equations the multipliers $\lambda^{\rho}$, $\mvec{\lambda}^p$,
  $\mvec{\lambda}^s$, and $\lambda^u$ can be calculated.
 Between the specific spin density and the orientation change
velocity $\mvec{\Omega}$ the relation \be \mvec{s} =
\mvec{\Theta}\cdot \mvec{\Omega} \en holds. $\mvec{\Theta}$ is the
moment of inertia tensor, which is assumed to be constant. In case
of suspensions of flexible fibers this is only an approximation,
because in principle this tensor changes if the fibers are
deformed. We assume here that these deformations are small and the
variation of $\mvec{\Theta}$ can be neglected. Then the specific
spin density $\mvec{s}$ depends only on the orientation change
velocity $\mvec{\Omega}$, and all other partial derivatives
vanish. In this case we have \be \frac{\partial \mvec{s}}{\partial
\mvec{\Omega}} = \mvec{\Theta} \en and the equations for the
multipliers simplify to  \beg
\lambda^{u} = - \frac{\frac{\partial \eta}{\partial T}}{\frac{\partial \e}{\partial T}}= -\frac{1}{T}\label{lu2} \\
\lambda^{\rho} = - \rho\frac{\partial \eta}{\partial \rho} + \frac{1}{T}\rho\frac{\partial \e}{\partial \rho}  \label{lo2} \\
\mvec{\lambda}^s = - \frac{\partial \left( \eta - \frac{1}{T}
\e\right) }{\partial \mvec{\Omega}}\cdot
\mvec{\Theta}^{-1}\label{ls2} \\
 \mvec{\lambda}^p = 0 \label{lp2} \quad . \ene for the derivative $\frac{\frac{\partial
\eta}{\partial T}}{\frac{\partial \e}{\partial T}}$ we insert
$\frac{1}{T}$. This relation is known in equilibrium. In
non-equilibrium it needs some additional argumentation, which can
be found in \cite{MULLER}.

The remaining equations (\ref{dot_nabla_v}), (\ref{nabla_omega}),
(\ref{nabla_rho}), (\ref{nabla_T}),  (\ref{nabla_nabla_v}), and
(\ref{nabla_O}) give after inserting the multipliers the
restrictions on constitutive functions. This will be shown here
only under the assumption of a constant moment of inertia tensor.
We have: \beg \frac{\partial\left( \eta - \frac{1}{T} \e
\right)}{\partial \nabla \mvec{v} } =0 \label{15} \\ \mvec{\Pi} =
-\frac{1}{T} \left( \rho \frac{\partial \left( \eta -\frac{1}{T}
\e\right) }{\partial \mvec{O}}+ \frac{\partial
\mvec{\Phi}}{\partial \mvec{\Omega}}\right. \nonumber \\ + \left.
\frac{\partial \left( \eta - \frac{1}{T}\e \right)  }{\partial
\mvec{\Omega}}\cdot \mvec{\Theta}^{-1}\cdot \frac{\partial
\mvec{\Pi}}{\partial
\mvec{\Omega}}\right) \label{16}\\
\frac{\partial \mvec{\Phi}}{\partial \rho} - \frac{1}{T}
\frac{\partial \mvec{q}}{\partial \rho} = - \frac{\partial \left(
\eta -\frac{1}{T} \e\right) }{\partial \mvec{\Omega}} \cdot
\mvec{\Theta}^{-1} \cdot \frac{\partial \mvec{\Pi}}{\partial \rho}
\label{17a}\\
\frac{\partial \mvec{\Phi}}{\partial T} - \frac{1}{T}
\frac{\partial \mvec{q}}{\partial T} = - \frac{\partial \left(
\eta -\frac{1}{T} \e\right) }{\partial \mvec{\Omega}}\cdot
\mvec{\Theta}^{-1}\cdot \frac{\partial \mvec{\Pi}}{\partial T}
\label{18}\\
\frac{\partial \mvec{\Phi}}{\partial \nabla \mvec{v}} -
\frac{1}{T} \frac{\partial \mvec{q}}{\partial \nabla \mvec{v}} = -
\frac{\partial \left( \eta -\frac{1}{T} \e\right) }{\partial
\mvec{\Omega}}\cdot \mvec{\Theta}^{-1}\cdot \frac{\partial
\mvec{\Pi}}{\partial \nabla \mvec{v}} \label{19}\\
 \frac{\partial
\mvec{\Phi}}{\partial \mvec{O}} - \frac{1}{T} \frac{\partial
\mvec{q}}{\partial \mvec{O}} = - \frac{\partial \left( \eta
-\frac{1}{T} \e\right) }{\partial \mvec{\Omega}}\cdot
\mvec{\Theta}^{-1}\cdot \frac{\partial \mvec{\Pi}}{\partial
\mvec{O}} \label{20} \quad . \ene Introducing the free energy
density $f := \e -T \eta$ and the difference $\mvec{k} =
\mvec{\Phi} - \frac{1}{T}\mvec{q}$ the resulting restrictions on
the constitutive functions can be written as: \beg \frac{\partial
f}{\partial \nabla \mvec{v} } =0
\label{f_biax}\\
\mvec{\Pi} = \frac{1}{T^2}  \rho \frac{\partial f}{\partial
\mvec{O}}+ \frac{1}{T} \frac{\partial \mvec{\Phi}}{\partial
\mvec{\Omega}} - \frac{1}{T^2} \frac{\partial f}{\partial
\mvec{\Omega}}\cdot \mvec{\Theta}^{-1}\cdot \frac{\partial
\mvec{\Pi}}{\partial \mvec{
\Omega}} \label{pi_biax}\\
\frac{\partial \mvec{k}}{\partial u_i} = \frac{1}{T}
\frac{\partial f}{\partial \mvec{\Omega}} \cdot \mvec{\Theta}^{-1}
\cdot \frac{\partial \mvec{\Pi}}{\partial u_i}  \quad \mbox{for}
\quad u_i \in \{ \rho ,T, \nabla \mvec{v},  \mvec{O} \}
\label{k_biax} \quad .\ene

Equation (\ref{k_biax}) shows that the difference $\mvec{k} =
\mvec{\Phi} - \frac{1}{T}\mvec{q}$ is surely non-zero, if the free
energy density depends on the orientation change velocity
($\frac{\partial f}{\partial \mvec{\Omega}} \neq 0 $), and if the
couple stresses $\mvec{\Pi}$ depend on any of the variables $\rho
,T$, $\nabla \mvec{v}$, $\mvec{O} $. In this case the very
frequently made constitutive assumption (see for instance
Irreversible Thermodynamics) of the entropy flux $\mvec{\Phi}$
being heat flux divided by temperature is not fulfilled.

Equation (\ref{pi_biax}) is a differential equation for the couple
stress. It reduces to an algebraic equation, if the free energy
density does not depend on $\mvec{\Omega}$. In this case the
couple stresses can be calculated as a derivative of the free
energy density \be \mvec{\Pi} = \frac{1}{T^2} \rho \frac{\partial
f}{\partial \mvec{O}}\quad . \en

\subsection{The entropy production}

The residual inequality is built up by all terms in equation
(\ref{Liu_biax}), which contain no higher derivatives, but only
state space functions. The constraint equation (\ref{constraint})
has been inserted, and $z = \frac{r}{T}$.  We will deal here only
with the case of a constant moment of inertia tensor. For the
multipliers equations (\ref{lu2}) to (\ref{lp2}) have to be
inserted: \beg \sigma = \rho \frac{\partial \eta }{\partial
\mvec{Q}}: \mvec{\dot{Q}} - \frac{\rho}{2} \frac{\partial \eta
}{\partial \mvec{O}}\cdot \nabla \mvec{v}\cdot \left( \nabla
\mvec{Q}\right) \cdot \mvec{Q}^T \cdot : \mvec{\epsilon} +
\frac{\partial \mvec{\Phi}}{\partial \mvec{Q}}\cdot :
\frac{\partial \mvec{Q}}{\partial \mvec{x}}\nonumber \\  +
\lambda^{\rho} \nabla \cdot \mvec{v} -\mvec{\lambda}^s \cdot
\left(\frac{\partial \mvec{\Pi}^T}{\partial \mvec{Q}}\cdot :
\frac{\partial \mvec{Q}}{\partial \mvec{x}} +\rho \mvec{m} -
\mvec{\epsilon}: \mvec{t}\right)\nonumber \\ +\lambda^u
 \left( \rho \frac{\partial \e }{\partial
\mvec{Q}}: \mvec{\dot{Q}} - \frac{\rho}{2} \frac{\partial \e
}{\partial \mvec{O}}\cdot \nabla \mvec{v}\cdot \left( \nabla
\mvec{Q}\right) \cdot \mvec{Q}^T \cdot : \mvec{\epsilon}   +
(\mvec{\epsilon} : \mvec{t}) \cdot \mvec{s} \right. \nonumber \\+
\left.
 \frac{\partial \mvec{q}}{\partial \mvec{Q}} \cdot :
 \frac{\partial \mvec{Q}}{\partial \mvec{x}} - \mvec{t} : (\nabla
 \mvec{v}) + \rho \mvec{f}\cdot \mvec{v}\right) \nonumber \ene
 \beg
= \rho \frac{\partial \eta }{\partial \mvec{Q}}: \mvec{\dot{Q}} -
\frac{\rho}{2} \frac{\partial \eta }{\partial \mvec{O}}\cdot
\nabla \mvec{v}\cdot \left( \nabla \mvec{Q}\right) \cdot
\mvec{Q}^T \cdot : \mvec{\epsilon} + \frac{\partial
\mvec{\Phi}}{\partial \mvec{Q}}\cdot : \frac{\partial
\mvec{Q}}{\partial \mvec{x}}\nonumber \\   +
\frac{1}{T}\rho\frac{\partial f}{\partial \rho} \nabla \cdot
\mvec{v} -\frac{1}{T} \frac{\partial f  }{\partial
\mvec{\Omega}}\cdot \mvec{\Theta}^{-1}\cdot \left(\frac{\partial
\mvec{\Pi}^T}{\partial \mvec{Q}}\cdot : \frac{\partial
\mvec{Q}}{\partial \mvec{x}} +\rho \mvec{g} - \mvec{\epsilon}:
\mvec{t}\right)\nonumber \\ - \frac{1}{T}
 \left(  \rho \frac{\partial \e }{\partial
\mvec{Q}}: \mvec{\dot{Q}} - \frac{\rho}{2} \frac{\partial \e
}{\partial \mvec{O}}\cdot \nabla \mvec{v}\cdot \left( \nabla
\mvec{Q}\right) \cdot \mvec{Q}^T \cdot : \mvec{\epsilon}
+(\mvec{\epsilon} : \mvec{t}) \cdot \mvec{s} \right. \nonumber \\
+
 \left. \frac{\partial \mvec{q}}{\partial \mvec{Q}} \cdot :
 \frac{\partial \mvec{Q}}{\partial \mvec{x}} - \mvec{t} : (\nabla
 \mvec{v}) + \rho \mvec{f}\cdot \mvec{v} \right)\nonumber  \ene
 \beg
 =\underbrace{- \frac{\rho}{T} \frac{\partial f }{\partial
\mvec{Q}}: \mvec{\dot{Q}}}_{\mbox{change of orientational
order}}\nonumber\\  + \underbrace{ \frac{\rho}{2T} \frac{\partial
f }{\partial \mvec{O}}\cdot \nabla \mvec{v}\cdot \left( \nabla
\mvec{Q}\right) \cdot \mvec{Q}^T \cdot :
\mvec{\epsilon}}_{\mbox{coupling between viscous flow and
orientational order}} \nonumber \\+ \underbrace{\frac{\partial
\mvec{k}}{\partial \mvec{Q}}\cdot : \frac{\partial
\mvec{Q}}{\partial \mvec{x}}}_{\mbox{transport of orientation}} +
\underbrace{ \frac{1}{T}\rho\frac{\partial f}{\partial \rho}
\nabla \cdot \mvec{v}}_{\mbox{viscous flow}}\nonumber\\
-\underbrace{\frac{1}{T} \frac{\partial f }{\partial
\mvec{\Omega}}\cdot \mvec{\Theta}^{-1}\cdot \left(\frac{\partial
\mvec{\Pi}^T}{\partial \mvec{Q}}\cdot : \frac{\partial
\mvec{Q}}{\partial \mvec{x}}\right)}_{\mbox{coupling between
change of orientation and gradient of orientation}} \nonumber
\\ \frac{1}{T} \frac{\partial f }{\partial
\mvec{\Omega}}\cdot \mvec{\Theta}^{-1}\cdot \left( \rho \mvec{m}
-\mvec{\epsilon} : \mvec{t} \right)\nonumber \\
 - \frac{1}{T}
 \left( (\mvec{\epsilon} : \mvec{t}) \cdot \mvec{s}  - \mvec{t} : (\nabla
 \mvec{v}) + \rho \mvec{f}\cdot \mvec{v} \right)\geq 0 \label{entropy_liu}
 \quad .\ene
The interpretation of the different terms is given in the
equation.

The first term in the entropy production, due to change of the
orientational order is analogous to the terms involving
$\alpha^0$, $\mvec{\alpha}^d$, and $\mvec{T}$ in the first
section. Both are due to changes of the internal variable. The
next term is new, because in this part here we have allowed for a
dependence of the free energy density on the spatial gradient of
the internal variable. A term due to transport of orientation does
not show up in the entropy production according to Irreversible
Thermodynamics because there the vector $\mvec{k}$ is assumed to
be always zero. The next term due to viscous flow and the term
$\mvec{t}:\nabla \mvec{v}$ appear in both method of exploiting the
dissipation inequality. Terms involving the derivative of the free
energy density with respect to the orientation change velocity are
new, because the rotation of fibers was not taken into account in
the Irreversible Thermodynamics treatment. On the other hand, in
this section we were not interested in heat conduction, and the
temperature gradient was not included in the state space.

\section{Mesoscopic theory of fiber suspensions}\label{meso}
\setcounter{equation}{0}

Fiber suspensions are an example of so called complex materials, meaning materials with an internal
structure, which can change under the action of external fields resulting in
 complex material behavior. The element of internal structure in our example is the orientation and
deformation of the fiber. There are two principally different
possibilities to deal with complex materials within continuum
thermodynamics: The first way is to introduce additional fields
depending on position and time. These fields can be internal
variables \cite{MU90,MAUMU94}, order or damage parameters
\cite{MAU92}, Cosserat-triads \cite{HANDIII3}, directors
\cite{ER60,LES65}, alignment and conformation tensors
\cite{Hess75,MMM}. The other way is a so called mesoscopic theory.
The idea is to enlarge the domain of the field quantities. The new
mesoscopic fields are defined on the space $\mathbb R^3_x\times
\mathbb R_t\times \M$. The manifold $\M$ is given by the set of
values the internal
 degree of freedom can take. Therefore the choice of $\M$ depends on the
 complex material under consideration. We will see later that the manifold $\M$ should be such that differentiation
and integration is possible on it. We call this way of dealing
with the internal structure of complex materials a mesoscopic
concept, because it includes more information than a macroscopic
theory on $\mathbb R^3_x\times \mathbb R_t$, but  the molecular
level is not considered like in a microscopic approach. The
mesoscopic level is between the microscopic and the macroscopic
level. The domain of the mesoscopic field quantities  $\mathbb
R^3_x\times \mathbb R_t\times \M$ is called mesoscopic space. The
orientation of an undeformed  fiber is described by a unit vector
$\bold{ \hat{ n}}$, where turning around the fiber by $\pi$ does
not change the orientation
 and therefore $  \bold{ \hat{ n}} \rightarrow -\bold{ \hat{ n}} $ is a symmetry transformation. The vector
 $\bold{ \hat{n}}$ is an element of the
unit sphere $S^2$. The deformation of the fiber is given by the
vector $\hat \varphi$ introduced in the first part. The difference
between that first part of the paper and the
 mesoscopic theory is, that  now fiber orientation and deformation are not introduced as macroscopic quantities,
 but as variables in the domain of fields. To distinguish these two kinds of quantities all mesoscopic
quantities are denoted with a $\hat{\vphantom{0.3cm}}\ $. We will
see later that this way we can deal with different fiber
orientations and deformations within one volume element of
continuum theory. In our example of fibers the manifold $\M$ is
the product of the set of values the orientation and deformation
 of a fiber can take: $\M = S^2 \times \mathbb R^3$ with $\mvec{\hat{n}} \in S^2$ and
 $\mvec{\hat \varphi} \in \mathbb R^3$.
 In order to be treated within a mesoscopic theory the length of
the fibers must be smaller than the linear
 dimension of the continuum element, meaning that the fibers must be smaller than any macroscopically
interesting length scale. Therefore this mesoscopic concept cannot be applied to fiber composite
materials with long fibers.

Beyond the use of additional variables $\m$ the mesoscopic concept
introduces a statistical element, the so-called {\em mesoscopic
distribution function} (MDF) generated by the different values of
the mesoscopic variable
 in a volume element. Here this is a distribution of  fiber orientations and deformations
$f(\bold {\hat{ n}}, \bold{ \hat {\varphi }}, \x , t )$ in the
volume element at position $\x$ and time t. This distribution
function gives the probability density of finding a fiber of a
specified orientation and deformation in this volume element. It
is normalized \be \int_{\mathbb R^3}\int_{S^2} f(\bold {\hat{
n}},\bold{ \hat {\varphi}} , \x,t)\, d^2\hat{ n} d^3\hat{ \bold{
\varphi}} =\ \mbox{$1$}. \quad . \en

\subsection{Mesoscopic balance equations}
\setcounter{equation}{0}

Now  fields as mass density, momentum density, etc. are defined on
the mesoscopic space. For distinguishing these fields from the
macroscopic ones we add the word ``mesoscopic". For instance the
mesoscopic mass density $\hat \varrho (\bold{ \hat{ n}}, \bold{
\hat{ \varphi}} , \x , t)$ is the mass density taking into account
only fibers of a specified orientation $\bold{ \hat{ n}}$ and
deformation $\bold{ \hat{ \varphi}}$. The macroscopic mass density
is the integral over all possible orientations and deformations:
\be \int_{\mathbb R^3}\int_{S^2} \hat{\varrho}(\bold{ \hat{
n}},\bold{ \hat{ \varphi}} , \x,t)\, d^2\hat n
d^3\hat{\bold{\varphi}} = \varrho (\x ,t) \quad . \en

From this equation and the interpretation of the distribution
function as probability density it is clear that the MDF  is given
by the mass fraction: \be f(\bold{ \hat{ n}},\bold{ \hat{
\varphi}} , \x,t)= \frac{\varrho(\bold{ \hat{ n}},\bold{ \hat{
\varphi}} , \x,t)}{\varrho (\x ,t)  } \quad . \en

For any kind of set of mesoscopic variables balance equations can
be derived for the mesoscopic field quantities (see for instance
\cite{Bam}) starting out from the macroscopic global ones
\C{PHYSICA, MCLC, Bam, mesocrack}. A generalized Reynolds
transport theorem in the mesoscopic  space \cite{Ehrentraut_diss}
is used to transform the time derivative, and Gauss theorem is
applied. For the fiber suspensions in regular points of the
continuum there result  the local mesoscopic balance equations. We
will show here only the balance of mass and the balance of
momentum with the abbreviation $(\cdot) = ( \bold{ \hat{ n}}
,\bold{ \hat{ \varphi}} ,\x ,t)$. Similarly mesoscopic balance
equations of energy and of angular momentum can be derived. There
is a balance of angular momentum independently  from the balance
 of momentum, because rotations of the fibers result in an internal angular momentum.

{\em Mesoscopic balance of mass} \be \frac{\partial}{\partial
t}\hat{\varrho}(\cdot)\ +\
  \nabla_{x}\cdot\left(\hat{\varrho}(\cdot)\hat{\mvec{v}}(\cdot
  )\right)
  +\nabla_{n}\cdot\left(\hat{\varrho}(\cdot)\mbox{\boldmath{$\dot{ \hat {n}}$}}(\cdot )\right) +
\nabla_{\hat{\varphi}}\left(  \varrho(\cdot)\dot {\hat {\bold
\varphi}}  (\cdot )\right)\ =\ 0. \en $\hat{\mvec{v}}(\cdot )$ is
the material velocity of fibers of a specified orientation and
length, $\dot {\hat {\bold n}}$ is the orientation change
velocity, and $ \dot {\hat {\bold \varphi}}(\cdot)$ is the
deformation change velocity.

{\em Mesoscopic balance of momentum}
\begin{eqnarray}
  \frac{\partial}{\partial
  t}\left(\hat{\varrho}(\cdot)\mbox{\boldmath{$\hat{v}$}}(\cdot)\right)+
\nabla_{x}\cdot\left(\mbox{\boldmath{$\hat{v}$}}(\cdot)\hat{\varrho}(\cdot)
  \mbox{\boldmath{$\hat{v}$}}(\cdot) - \mvec{\hat{t}}(\cdot)\right)\ +\
\nonumber\\ + \nabla_{n}\cdot\left(\bold {\hat {\dot  n}}(\x
,t)\hat{\varrho}(\cdot)
  \mbox{\boldmath{$\hat{v}$}}(\cdot) - \mvec{\hat{T}}(\cdot)\right)  + \nabla_{\varphi}
\left( \dot{\hat{ \varphi}}\hat{ \varrho}(\cdot)
\mbox{\boldmath{$\hat{v}$}}(\cdot) -
 \mvec{\hat{\tau}}(\cdot) \right)\  =\nonumber \\  \hat{\varrho}(\cdot)
 \mbox{\boldmath{$\hat{f}$}}(\cdot).
\label{mesostress}\end{eqnarray}

Here $\mvec{\hat{f}}(\cdot)$ is the external acceleration density,
$ \mbox{\boldmath{$\hat{t}$}}(\cdot)$ the  stress tensor, and $
\mvec{\hat{T}}(\cdot)$ the stress tensor on orientation space
(non-convective momentum flux in orientation space),
$\hat{\tau}(\cdot )$ is the momentum flux vector with respect to
the fiber deformation variable, all quantities defined on the
mesoscopic set of variables. \vspace{.3cm}\newline

As always in continuum theory the balance equations do not form a closed set of differential equations,
but constitutive equations are needed, now on the mesoscopic level. We are interested here mainly in
the constitutive equation for the stress tensor.

\subsection{Orientational order parameter and deformation variable}
The aim is to introduce macroscopic quantities from this
mesoscopic background, which describe the distribution of fiber
orientations and the average distortion of fibers. These are
internal variables in the sense of thermodynamics:

 Orientational
order parameters \be A^k = \int_{S^2}\int_{\mathbb R^3}
f(\mvec{\hat{\varphi}}, \mvec{\hat{n}}, \mvec{x}, t)
\underbrace{\mvec{\hat{n}}\dots \mvec{\hat{n}}}_k d^3\hat{\varphi}
d^2\hat{n} \quad ,\en deformation order parameters: \be \Phi^k
=\int_{S^2}\int_{\mathbb R^3} f(\mvec{\hat{\varphi}},
\mvec{\hat{n}}, \mvec{x}, t) \underbrace{\mvec{\hat{\varphi}}\dots
\mvec{\hat{\varphi}}}_k d^3\hat{\varphi}d^2\hat{n} \quad .\en

These order parameters are tensors of successive order. They are
macroscopic fields depending on position and time. With respect to
fiber orientations we have the symmetry transformation
$\mvec{\hat{n}} \rightarrow -\mvec{\hat{n}}$. Therefore all odd
order orientational order parameters vanish, and the first
non-zero order parameter, apart from the isotropic part $A^0 =1$
is the second order one: $\mvec{A}^2$. This can be included as an
internal variable in a macroscopic exploitation of the dissipation
inequality, like the tensor internal variable $\mvec{Q}$ in
section \ref{Liu}.

With the mesoscopic background it is possible to derive also equations of motion for these variables, which will be left
for a future work.

\subsection{Mesoscopic and macroscopic stress tensor}

The extensive quantity
 momentum has to be the integral over all mesoscopic momenta:
\be \rho (\x,t) \mvec{v} (\x,t)= \int_{S^2} \int_{\mathbb
R^3}\hat{\rho} (\x,t,\mvec{\hat{\varphi}}\ ,\mvec{\hat{n}} )
\mvec{\hat{v}} (\x,t,\mvec{\hat{\varphi}}\ ,\mvec{\hat{n}}  ) d^3
\hat{\varphi}d^2 \hat{n} \en

Integrating equation \R{mesostress} over all fiber orientations
and fiber deformations we must obtain the macroscopic balance of
momentum \R{eq:3.4}. This gives a relation between the mesoscopic
and the macroscopic stress tensor. With the abbreviation $(\cdot )
= (\x,t,\mvec{\hat{\varphi}}\ ,\mvec{\hat{n}}  )$ we have
\begin{eqnarray}
\int_{S^2} \int_{\mathbb R^3} \left( \frac{\partial}{\partial
  t}\left(\hat{\varrho}(\cdot)\mbox{\boldmath{$\hat{v}$}}(\cdot)\right)+
\nabla_{x}\cdot\left(\mbox{\boldmath{$\hat{v}$}}(\cdot)\hat{\varrho}(\cdot)
  \mbox{\boldmath{$\hat{v}$}}(\cdot) - \mvec{\hat{t}}(\cdot)\right)\right. \ +\
\nonumber\\\left.  + \nabla_{n}\cdot\left(\bold {\hat {\dot
n}}(\x ,t)\hat{\varrho}(\cdot)
  \mbox{\boldmath{$\hat{v}$}}(\cdot) - \mvec{\hat{T}}(\cdot)\right)  + \nabla_{\varphi}
\left( \dot{\hat{ \varphi}}\hat{ \varrho}(\cdot)
\mbox{\boldmath{$\hat{v}$}}(\cdot) -
 \mvec{\hat{\tau}}(\cdot) \right)d^3 \hat{\varphi}d^2
\hat{n}\right) \  =\nonumber \\ \int_{S^2} \int_{\mathbb R^3}
\hat{\varrho}(\cdot)\mbox{\boldmath{$\hat{f}$}}(\cdot)d^3
\hat{\varphi}d^2 \hat{n}\quad .
\end{eqnarray}
Integrating the divergence terms over the whole mesoscopic space
gives, according to GAUSS-theorem boundary terms. The
orientational part of the mesoscopic space, the unit sphere, is a
closed surface, i.e. without boundary. Concerning the deformation
variable we suppose that deformation cannot have arbitrarily large
values, and therefore there is no flux over the boundary at
infinity. We conclude that the last two integrals on the left hand
side vanish, and we end up with the equation
\begin{eqnarray}
\frac{\partial}{\partial
  t}\int_{S^2} \int_{\mathbb R^3}
\left(\hat{\varrho}(\cdot)\mbox{\boldmath{$\hat{v}$}}(\cdot)\right)d^3
\hat{\varphi}d^2 \hat{n} \nonumber \\ + \nabla_{x}\cdot
\int_{S^2} \int_{\mathbb R^3}
\left(\mbox{\boldmath{$\hat{v}$}}(\cdot)\hat{\varrho}(\cdot)
  \mbox{\boldmath{$\hat{v}$}}(\cdot) - \mvec{\hat{t}}(\cdot)\right)d^3
\hat{\varphi}d^2 \hat{n} \  =\nonumber
\\ \int_{S^2} \int_{\mathbb R^3}
\hat{\varrho}(\cdot)\mbox{\boldmath{$\hat{f}$}}(\cdot)d^3
\hat{\varphi}d^2 \hat{n} \quad .
\end{eqnarray}

Comparing the divergence terms under the space derivative with
those in the macroscopic balance of momentum  we have: \be
\int_{S^2} \int_{\mathbb R^3}\left(
\mbox{\boldmath{$\hat{v}$}}(\cdot)\hat{\varrho}(\cdot)
  \mbox{\boldmath{$\hat{v}$}}(\cdot) - \mvec{\hat{t}}(\cdot)\right) d^3
\hat{\varphi}d^2 \hat{n}=\rho \mvec{v}\mvec{v} - \mvec{t}\quad .
\en We conclude that for the constitutive quantity stress tensor
the following relation holds

\be \mvec{t}(\x , t) = \int_{S^2} \int_{\mathbb R^3}\left(
\mvec{\hat{t}}(\cdot)- \hat{\varrho}(\cdot)\mbox{\boldmath{$\delta
\hat{v}$}}(\cdot)
  \mbox{\boldmath{$\delta \hat{v}$}}(\cdot)\right) d^3
\hat{\varphi}d^2 \hat{n}   \en
 with the abbreviation \be
\delta \mvec{v} (\cdot )=\mbox{\boldmath{$\hat{v}$}}(\cdot)
-\mvec{v}(\mvec{x},t) \en

This result shows  that the fluxes, (here the stress tensor) are
in general not simply the integrals of the corresponding
mesoscopic quantities over the mesoscopic variables. This is true
only for the extensive quantities.

 If all fibers have the same translational velocity, then $\delta
\hat{\mvec{v}} (\cdot )=\mvec{0}$ and the macroscopic stress
tensor is the integral over all mesoscopic ones: \be \mvec{t}
=\int_{S^2} \int_{\mathbb R^3}\mvec{\hat{t}}(\cdot)d^3
\hat{\varphi}d^2 \hat{n} \quad . \label{t} \en

The mesoscopic stress tensor is a constitutive quantity, defined
on a suitable set of variables. This set of variables can include
mesoscopic quantities as well as macroscopic quantities. A
reasonable assumption for the set of variables is: \be \hat{Z} =
\lbrace \rho , T, \mvec{\hat{n}}, \mvec{\hat{\varphi}},
\karika{\mvec{d}}, \nabla \times \mvec{v} \rbrace \quad .\en

With this set of variables a representation theorem to linear
order in the velocity gradient and the deformation variable
$\hat{\varphi}$  gives the following expression for the mesoscopic
stress tensor: \beg \mvec{\hat{t}} = \frac{\hat{ \rho}}{\rho}
\left( \alpha_1 \mvec{\hat{n}}\mvec{\hat{n}} + \alpha_2
\mvec{\hat{n}} \mvec{\hat{\varphi}}+ \alpha_3
\mvec{\hat{\varphi}}\mvec{\hat{n}} + \alpha_4 \mvec{\hat{n}}
(\nabla \times \mvec{v})+ \alpha_5 (\nabla
\times \mvec{v})\mvec{\hat{n}} +\alpha_6 \karika{d}\right. \nonumber
\\\left.
+\alpha_7 \mvec{\hat{n}}  \mvec{\hat{n}} \cdot \karika{d}+\alpha_8
\mvec{\hat{n}} \cdot \karika{d}\mvec{\hat{n}}+\alpha_9
\mvec{\hat{n}} \cdot \karika{d}\cdot \mvec{\hat{n}} \mvec{\hat{n}}
\mvec{\hat{n}}\right)
 \quad ,\ene where
the material coefficients $\alpha_1$ to $\alpha_9$ may all depend
on the (macroscopic) mass density $\rho$ and temperature $T$.

Averaging this over the mesoscopic variables we obtain according
to equation (\ref{t}):

\beg \mvec{t} =\int_{S^2} \int_{\mathbb R^3} \frac{\hat{
\rho}}{\rho}\left(\alpha_1 \mvec{\hat{n}}\mvec{\hat{n}} + \alpha_2
\mvec{\hat{n}} \mvec{\hat{\varphi}}+ \alpha_3
\mvec{\hat{\varphi}}\mvec{\hat{n}} + \alpha_4 \mvec{\hat{n}}
(\nabla \times \mvec{v})+ \alpha_5 (\nabla
\times \mvec{v})\mvec{\hat{n}} \right. \nonumber \\
\left. +\alpha_6 \karika{d} +\alpha_7 \mvec{\hat{n}}
\mvec{\hat{n}} \cdot \karika{d}+\alpha_8 \mvec{\hat{n}} \cdot
\karika{d}\mvec{\hat{n}}+\alpha_9 \mvec{\hat{n}} \cdot
\karika{d}\cdot \mvec{\hat{n}} \mvec{\hat{n}}
\mvec{\hat{n}}\right) d^3 \hat{\varphi}d^2 \hat{n} \quad .
\nonumber
\\= \alpha_1 \mvec{A}^2+ \alpha_2 \langle \mvec{\hat{n}}
\mvec{\hat{\varphi}}\rangle + \alpha_3\langle
\mvec{\hat{\varphi}}\mvec{\hat{n}} \rangle +\alpha_6 \karika{d}
 d^3\hat{\varphi}d^2 \hat{n}\nonumber \\  +\alpha_7 \mvec{A}
\cdot \karika{d}+\alpha_8 \karika{d}\cdot \mvec{A}+\alpha_9
\karika{d}: \mvec{A^{(4)}}\quad. \label{te}\ene The average of
$\alpha_4 \mvec{\hat{n}} (\nabla \times \mvec{v})$ vanishes,
because $\int_{S^2} f \hat{\mvec{n}}d^2n =0$ due to the symmetry
$\mvec{\hat{n}}\leftrightarrow -\mvec{\hat{n}}$, analogously for
the term with $ \alpha_5$. The averages $\langle\mvec{\hat{n}}
\mvec{\hat{\varphi}}\rangle $ and $\langle
\mvec{\hat{\varphi}}\mvec{\hat{n}}\rangle $ are non-zero, because
they are even functions of $\mvec{\hat{n}}$: \be
\mvec{\hat{\varphi}}(-\mvec{\hat{n}}) = -
\mvec{\hat{\varphi}}(\mvec{\hat{n}}) \en (see the first section)
and therefore \be  -
\mvec{\hat{n}}\mvec{\hat{\varphi}}(-\mvec{\hat{n}}) =
\mvec{\hat{n}} \mvec{\hat{\varphi}}(\mvec{\hat{n}})\quad .  \en

The stress tensor (\ref{te}) clearly can have an antisymmetric
part $\mvec{t}^a$: \be \mvec{t}^a =\frac{1}{2} \left( \alpha_2
-\alpha_3 \right) \left( \langle \mvec{\hat{n}}
\mvec{\hat{\varphi}}\rangle  - \langle
\mvec{\hat{\varphi}}\mvec{\hat{n}} \rangle\right)+ \frac{1}{2}
\left( \alpha_7 -\alpha_8 \right) \left( \mvec{A} \cdot
\karika{d}- \karika{d}\cdot \mvec{A}\right)  \quad .  \en

\section{Conclusions}
\setcounter{equation}{0}

We have investigated the constitutive properties of suspensions of
long fibers, which can be deformed. In the second and in the third
part we have taken into consideration in addition the possibility
of different fiber orientations. This complex material has been
treated with three different methods. Two of these methods are
different ways of exploiting the second law of thermodynamics in
macroscopic continuum  thermodynamics. The first method was
classical Thermodynamics of Irreversible Processes (TIP). The
second method of exploitation, the method according to LIU, is
more related to Rational Thermodynamics. In both methods the aim
is a derivation of information on constitutive functions, and the
first  method shows that in fiber suspensions there can exist an
antisymmetric part of the stress tensor due to internal degrees of
freedom. The method of TIP gives additional relaxation equations
for the internal variables introduced in the beginning in order to
account for the internal structure. Such relaxation equations
cannot be derived by the method of LIU. The assumptions made in
the beginning by this method are less restrictive, and the
constitutive equations consistent with the results of the method
by LIU are more general than the equations derived from TIP. It
could be shown that the assumption made in TIP concerning the
entropy flux being heat flux divided by temperature does not
contradict the results of the LIU-procedure, if  the free energy
density does not depend on the orientation change velocity of the
fibers. This assumption was made by the choice of variables in the
first section. Therefore this exploitation is in agreement with
the results of the LIU-procedure. However, if the orientation
change velocity is included in the set of variables, the entropy
flux might be not simply heat flux divided by temperature, as a
result of the exploitation of the dissipation inequality. This can
be interpreted as non-convective entropy transport due to change
of fiber orientations.  In addition couple stress could be
calculated as a derivative of the free energy.

Finally the idea of a so called mesoscopic theory was sketched.
There we introduced a finer description considering single fibers
with different orientations and different states of deformation.
The macroscopic quantities of the usual continuum theory are
obtained by averaging over the different fiber orientations and
different states of  deformation. This method gives, apart from
the usual balance equations, a definition of internal variables in
terms of averages. Equations of motion for these internal
variables will be discussed in a future paper. Another result
shown here is the fact that the stress tensor calculated as an
average of mesoscopic quantities has an antisymmetric part.

In the mesoscopic theory we did not deal further with the
possibility that the orientation distribution of fibers and also
the distribution of fiber deformations can change, for instance
under the action of a flow field. It is expected that both effects
have influence on
 material properties, a problem, that will be dealt with in a future work.

\section*{Acknowledgements}
This research was supported by OTKA T034715 and T034603. We thank
the ''Deutsche Akademische Austauschdienst'' and the ''Deutsche
Forschungsgemeinschaft'' for sponsoring the cooperation between
both the Departments of Physics and Chemical Physics  of the
Technical University of Berlin and the Budapest University of
Technology and Economics. Financial support by the VISHAY Company,
95100 Selb, Germany, is gratefully acknowledged.

\newpage

{\bf \large Appendix: Proof of the proposition on the time
derivative of the state space variable $\mvec{O}$}

 \vspace*{0.5cm}

{\bf Proposition}:

\be \frac{d \mvec{O}}{dt}= \left( \nabla \mvec{\Omega} \right)^T +
\frac{1}{2} \left( \mvec{\epsilon} \cdot \mvec{Q} : \frac{\partial
\mvec{Q}^T}{\partial \mvec{x}}\right)\cdot \frac{\partial
\mvec{v}}{\partial \mvec{x}} \en or in components \be \frac{d
O_{ik}}{dt}=  \frac{\partial \Omega_i}{\partial x_k}  +
\frac{1}{2} \epsilon_{irl} Q_{ls}  \frac{\partial Q_{rs}}{\partial
x_m} \frac{\partial v_m}{\partial x_k} \quad . \en Proof:\newline
 In components we have \beg \frac{d O_{ik}}{dt}=  -
\frac{1}{2} \epsilon_{irl} \frac{d }{dt}\left( \frac{\partial
Q_{rs}}{\partial x_k} Q_{ls} \right)
= \nonumber \\
 - \frac{1}{2} \epsilon_{irl}\left(  \frac{d }{dt}\left( \frac{\partial Q_{rs}}{\partial x_k}\right)  Q_{ls}+
  \frac{\partial Q_{rs}}{\partial x_k}\frac{d Q_{ls} }{dt}\ \right) = \nonumber \\
= - \frac{1}{2} \epsilon_{irl}\left(  \left( \frac{\partial
}{\partial t} + v_m \frac{\partial }{\partial x_m} \right)
 \left( \frac{\partial Q_{rs}}{\partial x_k}\right)  \right. Q_{ls}+\nonumber
 \\\left.
   \frac{\partial Q_{rs}}{\partial x_k}\epsilon_{lkm}\Omega_k Q_{ms} \right) = \nonumber \\
=- \frac{1}{2} \epsilon_{irl}\left(  \left( \frac{\partial
}{\partial x_k} \left( \frac{\partial Q_{rs}}{\partial t} + v_m
\frac{\partial Q_{rs} }{\partial x_m} \right)-  \frac{\partial
v_m}{\partial x_k} \frac{\partial Q_{rs}}{\partial x_m}\right)
 Q_{ls}\right. +\nonumber \\
  \left.  \frac{\partial Q_{rs}}{\partial x_k}\epsilon_{lkm}\Omega_k Q_{ms} \right) = \nonumber \\
= - \frac{1}{2} \epsilon_{irl}\left(  \left( \frac{\partial
}{\partial x_k} \frac{d Q_{rs}}{d t}- \frac{\partial v_m}{\partial
x_k} \frac{\partial Q_{rs}}{\partial x_m}\right)
 Q_{ls}\right. +\nonumber \\
  \left.  \frac{\partial Q_{rs}}{\partial x_k}\epsilon_{lkm}\Omega_k Q_{ms} \right) = \nonumber \\
= - \frac{1}{2} \epsilon_{irl}\left(  \left( \frac{\partial
}{\partial x_k} \left( \epsilon_{rop} \Omega_o Q_{ps}\right)-
\frac{\partial v_m}{\partial x_k} \frac{\partial Q_{rs}}{\partial
x_m}\right)
 Q_{ls}\right. +\nonumber \\
  \left.  \frac{\partial Q_{rs}}{\partial x_k}\epsilon_{lkm}\Omega_k Q_{ms} \right) = \nonumber \\
 \frac{1}{2} \left( \left( \delta_{io} \delta_{lp} -
 \delta_{ip}\delta_{lo}\right)\left( \frac{\partial \Omega_o}{\partial
 x_k} Q_{ps} + \Omega_o \frac{\partial Q_{ps}}{\partial x_k}
 \right) +\frac{1}{2} \epsilon_{irl}\frac{\partial v_m}{\partial
 x_k} \frac{\partial Q_{rs}}{\partial x_m} \right) Q_{ls}
 \nonumber\\ -  \frac{1}{2} \left( \delta_{ik} \delta_{rm} -
 \delta_{im}\delta_{rk}\right)\frac{\partial Q_{rs}}{\partial x_k}
 \Omega_k Q_{ms} =\nonumber \\
=\frac{1}{2} \left(  \frac{\partial \Omega_i }{\partial
x_k}Q_{ls}Q_{ls} - \frac{\partial \Omega_l }{\partial
x_k}Q_{is}Q_{ls}+ \epsilon_{irl}\frac{\partial v_m}{\partial
 x_k} \frac{\partial Q_{rs}}{\partial x_m}  Q_{ls}\right)
= \nonumber \\
=\frac{\partial \Omega_i }{\partial x_k}
+\frac{1}{2}\epsilon_{irl}\frac{\partial v_m}{\partial
 x_k} \frac{\partial Q_{rs}}{\partial x_m}  Q_{ls}
\quad.  \ene



\begin{thebibliography}{10}

\bibitem{Lhullier}
D.~Lhullier.
\newblock Phenomenology of hydrodynamic interactions in suspensions of weakly
  deformable particles.
\newblock {\em J. Physique}, 48:1887--1893, 1987.

\bibitem{Goddard}
J.~D. Goddard and C.~Miller.
\newblock {\em J. Fluid Mech.}, 28:657, 1967.

\bibitem{Roscoe}
R.Roscoe.
\newblock {\em J. Fluid Mech.}, 1967.

\bibitem{Schowalter}
W.~R. Schowalter, C.~E. Chaffrey, and H.~Brenner.
\newblock {\em J. Colloid. Interface Sci.}, 26:152, 1968.

\bibitem{Frankel}
N.~A. Frankel and A.~Acrivos.
\newblock {\em J. Fluid Mech.}, 44:65, 1970.

\bibitem{Lhullier_CISM}
D.~Lhullier.
\newblock Cism courses and lectures 370.
\newblock In U.~Schaflinger, editor, {\em The flow of particles in suspension}.
  Springer, 1996.

\bibitem{Shen}
W.~Shen, L.~Peng, and Y.~Yue.
\newblock Damage dependent viscoelastic constitutive relations for glass-fiber
  woven/polyester composite plate.
\newblock {\em Engineering Fracture Mechanics}, 47:867--872, 1994.

\bibitem{Lou}
Y.~C. Lou and R.~A. Schapery.
\newblock Viscoelastic characterization of nonlinear fiber-reinforced plastics.
\newblock {\em J. Compos.}, 5:208--234, 1971.

\bibitem{Schapery}
R.~A. Schapery.
\newblock Advances in aerospace structures and materials.
\newblock In S.~S.~Wand et. al., editor, {\em On viscolelastic deformation and
  failure behavior of composites with distributed flaws}, pages 387--398. ASME,
  1981.

\bibitem{Schapery2}
R.~A. Schapery.
\newblock A micromechanical model for linear viscoelastic behavior of
  particle-reinforced rubber with distributed damage.
\newblock {\em Engn. Fracture Mech.}, 25:845--867, 1986.

\bibitem{Doi_Edwards}
M.~Doi and S.~F. Edwards.
\newblock {\em The theory of polymer dynamics}.
\newblock Clarendon, Oxford, 1986.

\bibitem{Batchelor}
G.~K. Batchelor.
\newblock {\em J. Fluid Mech.}, 41:545--570, 1970.

\bibitem{Batchelor2}
G.~K. Batchelor.
\newblock {\em J. Fluid Mech.}, 46:813--829, 1971.

\bibitem{Hinch}
R.~A. Keiller and E.~J. Hinch.
\newblock Corner flow of suspensions of rigid rods.
\newblock {\em Journal of Non-Newtonian Fluid Mechanics}, 40:323--335, 1991.

\bibitem{Verhas2}
C.~Papenfuss, J.~Verhas, and W.~Muschik.
\newblock A simplified thermodynamic theory for biaxial nematics.
\newblock {\em Z. Naturforsch.}, 50a:795--804, 1994.

\bibitem{Rice}
J.~R. Rice.
\newblock Inelastic constitutive relations for solids: an internal-variable
  theory and its application to metal plasticity.
\newblock {\em J. Mech. Phys. Solids}, 19:433--455, 1971.

\bibitem{rat}
T.~Alts.
\newblock Thermodynamics of thermoelastic bodies with kinematic constraints.
  fibre reinforced materials.
\newblock {\em Arch. Rat. Mech. Anal.}, 61:253--289, 1976.

\bibitem{Liu}
I.~Shih Liu.
\newblock Method of {L}agrange multipliers for exploitation of the entropy
  principle.
\newblock {\em Arch. Rat. Mech. Anal.}, 46:131--148, 1972.

\bibitem{BLENK91}
S.~Blenk and W.~Muschik.
\newblock Orientational balances for nematic liquid crystals.
\newblock {\em J. Non-Equilib. Thermodyn.}, 16:67--87, 1991.

\bibitem{Journal}
H.~Ehrentraut, W.~Muschik, and C.~Papenfuss.
\newblock Mesoscopically derived orientation dynamics of liquid crystals.
\newblock {\em J. Non-Equilib. Thermodyn.}, 22:285--298, 1997.

\bibitem{ZAMM92}
S.~Blenk and W.~Muschik.
\newblock Mesoscopic concepts for constitutive equations of nematic liquid
  crystals in alignment tensor formulation.
\newblock {\em ZAMM}, 73(4-5):T331--T333, 1993.

\bibitem{PHYSICA}
S.~Blenk, H.~Ehrentraut, and W.~Muschik.
\newblock Statistical foundation of macroscopic balances for liquid crystals in
  alignment tensor formulation.
\newblock {\em Physica A}, 174:119--138, 1991.

\bibitem{Bam}
C.~Papenfuss.
\newblock Theory of liquid crystals as an example of mesoscopic continuum
  mechanics.
\newblock {\em Computational Materials Science}, 19:45 -- 52, 2000.

\bibitem{poz}
C.~Papenfuss and W.~Muschik.
\newblock Liquid crystal theory as an example of mesoscopic continuum
  mechanics.
\newblock In B.~T. Maruszewski, W.~Muschik, and A.~Radowicz, editors, {\em
  Trends in Continuum Physics}. World Scientific, 1998.

\bibitem{20}
G.~A. Kluitenberg.
\newblock On dielectric and magnetic relaxation phenomena and non-equilibrium
  thermodynamics.
\newblock {\em Physica}, 68:75--92, 1973.

\bibitem{21}
G.~A. Kluitenberg.
\newblock On dielectric and magnetic relaxation phenomena and vectorial
  internal degrees of freedom in thermodynamics.
\newblock {\em Physica}, 87A:302--330, 1977.

\bibitem{22}
G.~A. Kluitenberg.
\newblock On vectorial internal variables and dielectric and magnetic
  relaxation phenomena.
\newblock {\em Physica}, 109A:91--122, 1981.

\bibitem{23}
G.~A. Kluitenberg and V.~Ciancio.
\newblock On linear dynamical equations of state for isotropic media. i.
  general formalism.
\newblock {\em Physica}, 93A:273--286, 1978.

\bibitem{3}
V.~Ciancio and J.~Verh{\'a}s.
\newblock On heat conduction in media with isotropic microstructure.
\newblock {\em Atti.\ Academia Peloritania dei Pericolanti (Messina)},
  68:41--53, 1990.

\bibitem{4}
V.~Ciancio and J.~Verh{\'a}s.
\newblock A thermodynamic theory for radiating heat transfer.
\newblock {\em J.\ Non-Equilib.\ Thermodyn.}, 17:33--43, 1990.

\bibitem{5}
V.~Ciancio and J.~Verh{\'a}s.
\newblock On thermal wawes and radiating heat transfer.
\newblock {\em Acta Physica Hungarica}, 69:69, 1990.

\bibitem{6}
D.~Fekete.
\newblock A systematic application of gyarmati's wave theory of thermodynamics
  to thermal waves in solids.
\newblock {\em Phys.\ Stat.\ Sol.\ (b)}, 105:161--174, 1981.

\bibitem{7}
D.~Fekete.
\newblock Application of the fundamental principle of dissipative processes to
  gyarmati's wave theory. (in russian).
\newblock {\em Russ.\ J.\ Phys.\ Chem.}, 57:2700--2703, 1983.

\bibitem{8}
L.~S. Garcia-Colin and R.~F. Rodriguez.
\newblock On the relationship between extended thermodynamics and the wave
  approach to thermodynamics.
\newblock {\em J.\ Non-Equilib.\ Thermodyn.}, 13:81--94, 1988.

\bibitem{1}
Vincenzo Ciancio and G.~A. Kluitenberg.
\newblock On linear dynamical equations of state for isotropic media. ii. some
  cases of special interest.
\newblock {\em Physica}, 99A:592--600, 1979.

\bibitem{2}
V.~Ciancio, E.~Turrisi, and G.~A. Kluitenberg.
\newblock On the propagation of linear longitudinal acoustic waves in isotropic
  media with shear and volume viscosity and a tensorial internal variable. i.
  general formalism.
\newblock {\em Physica}, 125A:640--652, 1984.

\bibitem{11}
S.~R. De~Groot.
\newblock {\em Thermodynamics of Irreversible Processes.}
\newblock North-Holland Publ.\ Co., Amsterdam, 1951.

\bibitem{12}
S.~R. De~Groot and P.~Mazur.
\newblock {\em Non-equilibrium Thermodynamics.}
\newblock North-Holland Publ.\ Co., Amsterdam, 1962.

\bibitem{13}
I.~Gyarmati.
\newblock {\em Non-Equilibrium Thermodynamics.}
\newblock Springer, Berlin, 1970.

\bibitem{16}
D.~Jou, J.~M. Rubi, and J.~Casas-Vazquez.
\newblock A generalized gibbs equation for second order fluids.
\newblock {\em J.\ Phys.\ A}, 12:2515--2520, 1979.

\bibitem{18}
G.~A. Kluitenberg.
\newblock On the thermodynamics of viscosity and plasticity.
\newblock {\em Physica}, 29:633--652, 1963.

\bibitem{19}
G.~A. Kluitenberg.
\newblock On heat dissipation due to irreversible mechanical phenomena in
  continuous media.
\newblock {\em Physica}, 35:177--192, 1967.

\bibitem{24}
G.~A. Kluitenberg, E.~Turrisi, and V.~Ciancio.
\newblock On the propagation of linear transverse acoustic waves in isotropic
  media with mechanical relaxation phenomena due to viscosity and a tensorial
  internal variable. i. general formalism.
\newblock {\em Physica}, 110A:361--372, 1982.

\bibitem{29}
I.~M{\"u}ller.
\newblock {\em Thermodynamics.}
\newblock Pitman Publ.\ Co., London, 1985.

\bibitem{31}
B.~Ny{\'\i}ri.
\newblock Thermodynamical derivation of equations of motion for multicomponent
  fluids.
\newblock {\em Acta Phys.\ Hung.}, 60:245--260, 1986.

\bibitem{41}
E.~Turrisi, V.~Ciancio, and G.~A. Kluitenberg.
\newblock On the propagation of linear transverse acoustic waves in isotropic
  media with mechanical relaxation phenomena due to viscosity and a tensorial
  internal variable. ii. some cases of special interest (poynting-thomson,
  jeffrey, maxwell, kelvin-voigt, hooke and newton media).
\newblock {\em Physica}, 116A:594--603, 1982.

\bibitem{42}
K.~C. Valanis.
\newblock {\em Irreversible Thermodynamics of Continuous Media.}
\newblock Springer, Wien, 1971.

\bibitem{44}
J.~Verh{\'a}s.
\newblock A thermodynamic approach to viscoelasticity and plasticity.
\newblock {\em Acta Mechanica}, 53:125--139, 1984.

\bibitem{46}
J.~Verh{\'a}s.
\newblock Irreversible thermodynamics for the rheological properties of
  colloids.
\newblock {\em Int.\ J.\ Heat Mass Transfer}, 30:1001--1006, 1987.

\bibitem{45}
J.~Verh{\'a}s.
\newblock Irreversible thermodynamics of nematic liquid crystals.
\newblock {\em Acta Phys.\ Hung.}, 55:275--291, 1984.

\bibitem{J.V.Borkow-paper}
J.~Verh\'as.
\newblock Thermodynamic theory for couple stress.
\newblock In {\em Second Workshop on Dissipation in Physical Systems, 1997,
  Borkow, Poland}, To appear.

\bibitem{40}
V.~K. Stokes.
\newblock {\em Theory of Fluids with Microstructure.}
\newblock Springer, Berlin, 1984.

\bibitem{47}
J.~Verh{\'a}s.
\newblock {\em Thermodynamics and Rheology.}
\newblock Kluwer and Akad\'emiai Kiad\'o, Dordrecht and Budapest, 1997.

\bibitem{9}
C.~Garrod and J.~Hurley.
\newblock Symmetry relations for the conductivity tensor.
\newblock {\em Phys.\ Rev}, A 27:1487--1490, 1983.

\bibitem{10}
P.~Glansdorff and I.~Prigogine.
\newblock {\em Themodynamic Theory of Structure, stability and Fluctuations.}
\newblock Wiley--Interscience, London, 1971.

\bibitem{15}
J.~Hurley and C.~Garrod.
\newblock Generalization of the onsager reciprocity theorem.
\newblock {\em Phys.\ Rev.\ Lett.}, 48:1575--1577, 1982.

\bibitem{25}
S.~Machlup and L.~Onsager.
\newblock Fluctuations and irreversible processes ii. systems with kinetic
  energy.
\newblock {\em Phys.\ Rev.}, 91:1512--1515, 1953.

\bibitem{32}
L.~Onsager.
\newblock Reciprocal relations in irreversible processes i.
\newblock {\em Phys.\ Rev.}, 37:405--426, 1931.

\bibitem{33}
L.~Onsager.
\newblock Reciprocal relations in irreversible processes ii.
\newblock {\em Phys.\ Rev.}, 38:2265--2279, 1931.

\bibitem{34}
L.~Onsager and S.~Maclup.
\newblock Fluctuations and irreversible processes.
\newblock {\em Phys.\ Rev.}, 91:1505--1512, 1953.

\bibitem{35}
I.~Prigogine.
\newblock {\em Introduction to Thermodynamics of Irreversible Processes.}
\newblock Interscience, New York, 1961.

\bibitem{39}
G.~F. Smith and R.~S. Rivlin.
\newblock The anisotropic tensors.
\newblock {\em Quart.\ Appl.\ Math.}, 15:308--314, 1957.

\bibitem{amendment}
W.~Muschik.
\newblock An amendment to the second law of thermodynamics.
\newblock {\em J. Non-Equilib. Thermodyn.}, 21:175--192, 1996.

\bibitem{MULLER}
I.~Mueller.
\newblock {\em Thermodynamics}.
\newblock Pitman Advanced Publishing Program, Boston, London, Melbourne, 1985.

\bibitem{MU90}
W.~MUSCHIK.
\newblock Internal variables in non-equilibrium thermodynamics.
\newblock {\em J. Non-Equilib. Thermodyn.}, 15, 1990.

\bibitem{MAUMU94}
G.A. MAUGIN and W.~MUSCHIK.
\newblock Thermodynamics with internal variables.
\newblock {\em J. Non-Equilib. Thermodyn.}, 19, 1994.

\bibitem{MAU92}
G.A. MAUGIN.
\newblock {\em The Thermomechanics of Plasticity and Fracture, Chap. 7}.
\newblock Cambridge University Press, Cambridge, 1992.

\bibitem{HANDIII3}
C.~TRUESDELL and W.~Noll.
\newblock {\em Non-Linear Field Theories of Mechanics, Encyclopedia of Physics,
  Vol. III/3, Sect. 98}.
\newblock Springer Verlag, Berlin, etc., 1965.

\bibitem{ER60}
J.L. ERICKSEN.
\newblock Anisotropic fluids.
\newblock {\em Arch. Rat. Mech. Anal.}, 4, 1960.

\bibitem{LES65}
F.J. LESLIE.
\newblock Some constitutive equations for liquid crystals.
\newblock {\em Arch. Rat. Mech. Anal.}, 28, 1965.

\bibitem{Hess75}
S.~Hess.
\newblock Irreversible thermodynamics of nonequilibrium alignment phenomena in
  molecular liquids and in liquid crystals.
\newblock {\em Z. Naturforsch.}, 30a:728--733, 1975.

\bibitem{MMM}
G.A. MAUGIN and R.~DROUOT.
\newblock Thermodynamic modelling of polymers in solution.
\newblock In Axelrad and W.~Muschik, editors, {\em Constitutive Laws and
  Microstructure}. Springer Verlag, 1988.

\bibitem{MCLC}
S.~Blenk, H.~Ehrentraut, and W.~Muschik.
\newblock Orientation balances for liquid crystals and their representation by
  alignment tensors.
\newblock {\em Mol. Cryst. Liqu. Cryst.}, 204:133--141, 1991.

\bibitem{mesocrack}
P.~Van, C.~Papenfuss, and W.~Muschik.
\newblock Mesoscopic dynamics of microcracks.
\newblock {\em Physical Review E}, 62(5):6206--6215, 2000.

\bibitem{Ehrentraut_diss}
H.~Ehrentraut.
\newblock {\em A {U}nified {M}esoscopic {C}ontinuum {T}heory of {U}niaxial and
  {B}iaxial {L}iquid {C}rystals}.
\newblock Wissenschaft und Technik Verlag, Berlin, 1996.

\end{thebibliography}

\vspace{2em}

{\em E-mail addresses:} cpapenfuss@c8m42.pi.tu-berlin.de and
 Verhas@phy.bme.hu

\end{document}